\documentclass[12pt]{article}

\usepackage{newtxtext,newtxmath}
\usepackage{graphicx}

\usepackage[letterpaper,margin=1in]{geometry}

\renewenvironment{abstract}
	{\quotation}
	{\endquotation}

\date{}

\makeatletter
\renewcommand{\fnum@figure}{\textbf{Figure \thefigure}}
\renewcommand{\fnum@table}{\textbf{Table \thetable}}
\makeatother

\usepackage{scicite}

\usepackage{url}

\def\scititle{
	High-$Q$ microresonators unveil quantum rare events
}
\title{\bfseries \boldmath \scititle}

\author{
	Sricharan Raghavan-Chitra$^{1}$,
	Arghadip Koner$^{1\ast}$,
	Joel Yuen-Zhou$^{1\ast}$\and
	\small$^{1}$ Department of Chemistry and Biochemistry, University of California San Diego, La Jolla, California 92093\and
	\small$^\ast$Corresponding author. Email: akoner@ucsd.edu, joelyuen@ucsd.edu\and
}

\begin{document} 

% Insert the title and author list
\maketitle

% Abstract, in bold
% There are strict length limits, and not all formats have abstracts.
% Consult the journal instructions to authors for details.
% Do not cite any references in the abstract.
\begin{abstract} \bfseries \boldmath
Classical linear optics posits that at sufficiently low intensities, light propagation in dielectric media is governed solely by their linear susceptibilities. Here, we demonstrate a departure from this paradigm in high-$Q$ microresonators, where prolonged photon confinement enables rare quantum electrodynamical (QED) events—mediated by the quantum vacuum—to embed distinctive Raman signatures of the coupled analyte into the resonator’s linear transmission spectrum despite their absence from the linear susceptibility. We further show that increasing the amount of adsorbed analyte sample amplifies these Raman fingerprints well above typical noise floors, rendering them experimentally accessible with state-of-the-art photonic architectures and detection schemes. This novel weak-coupling cavity QED effect offers unique routes to harness extended photon lifetimes and constrained geometries for leveraging vacuum fluctuations in next-generation photonic technologies for chemical and biological sensing and high-precision optical spectroscopy.
\end{abstract}

% The first paragraph of any Science paper does NOT have a heading
% Nor is it indented
\noindent
\section{\label{sec:level1}Introduction}

\begin{figure*}[ht!]
        \includegraphics[width=\linewidth]{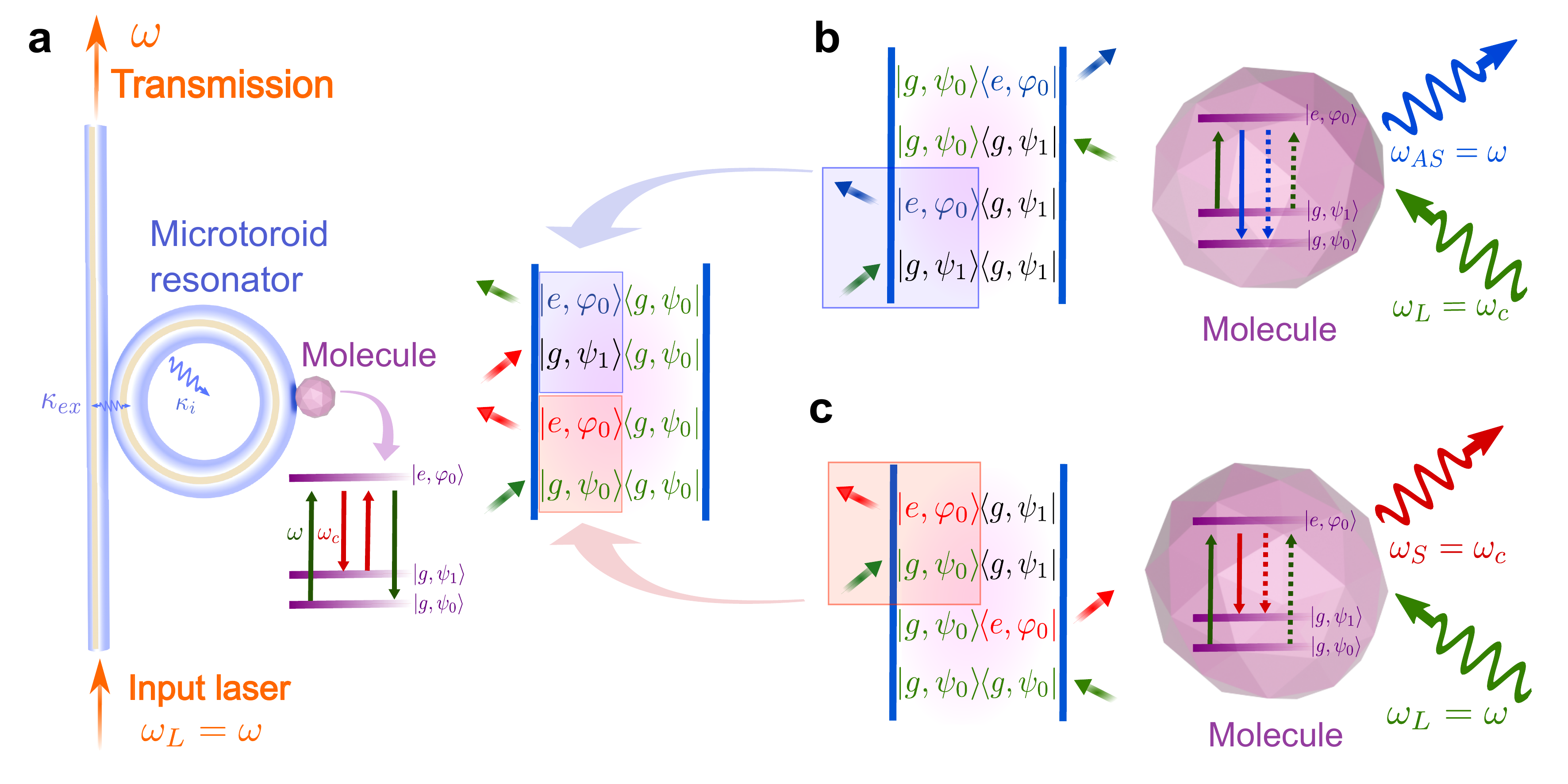}
\caption{
\textbf{Schematic illustration of vacuum-mediated Raman processes in the linear optics of a microresonator system, highlighting the Stokes and anti-Stokes components.} \small
    \textbf{a}, A high-$Q$ microtoroid resonator at frequency $\omega_c$ is evanescently coupled to a single molecule and probed with an incident laser at frequency $\omega$, introduced via an optical fiber through evanescent coupling. The key phenomenon we reveal is that contrary to classical linear optics predictions, a molecule’s Raman vibrational fingerprints directly manifest in \textit{linear} transmission, $T(\omega)$, of the microtoroid, provided the cavity lifetime is sufficiently long. The adjacent ladder and double-sided Feynman diagrams (DSFDs) illustrate a cavity Raman process, where Stokes scattering populates the cavity vacuum with a field that subsequently drives the anti-Stokes transition. Notably, this mechanism solely involves quantum coherences, unlike the Purcell effect. Moreover, distinct from conventional Raman, it \textbf{neither} induces vibrational heating or cooling \textbf{nor} results in the up- or downconversion of photons. Instead, the process operates within the linear regime, introducing additional absorption channels at new frequencies. This vacuum-mediated process cannot occur outside the single-mode cavity, as the Stokes field becomes irreversibly dispersed among the numerous modes of the free-field electromagnetic continuum. In contrast, \textbf{b}, (\textbf{c},)  denote a conventional anti-Stokes (Stokes) Raman process that occurs when an incident laser photon at frequency $\omega_L$ inelastically scatters to $\omega_{AS}$ ($\omega_S$) by extracting (leaving behind) vibrational energy from the molecule. This leads to vibrational cooling (heating) as the population is transferred to the vibrational ground (excited) state. Being an inherently nonlinear process, this upconvert (downconvert) the laser photon, unlike in the linear regime of the microtoroid resonator. The corresponding DSFD~\cite{MukamelBook} and ladder diagrams~\cite{tokmakoff_nonlinear_notes} illustrate the underlying mechanism. Solid and dotted lines in the ladder diagrams indicate the ket- and bra-side interactions in the DSFDs.}
\label{fig: schematic_raman_process}
\end{figure*}

\noindent Propagation of weak-intensity light through dielectric media is typically considered a classical optics affair~\cite{feynman2012electromagnetism,james1992why,jackson1998classical}. The electric field amplitudes of transmitted, reflected, and absorbed light can be accurately described using the standard framework based on the linear permittivity of the material, \(\epsilon^{(1)}(\boldsymbol{r},\omega) = \epsilon_0 [1 + \chi^{(1)}(\boldsymbol{r},\omega)]\), where \(\chi^{(1)}(\boldsymbol{r},\omega)\) represents its linear susceptibility and \(\epsilon_0\) is the permittivity of free space~\cite{saleh1991fundamentals,yariv2006photonics}. For example, light propagation through a multilayered dielectric medium can be accurately described using transfer matrix methods~\cite{nikolett2020transfer}. One may, of course, decide to care about the photon statistics of light, upon which the correct formalism is quantum linear optics, where linear permittivities are still the only needed material input~\cite{glauber1991quantum,monticone2014beating}. In this article, we investigate a novel light phenomenon that does not rely on measurements of quantum statistics or entanglement of the photons~\cite{Moradi2025photon,anderson2018two,england2016phonon,measuring2008waldermann}, yet the conventional framework of classical linear optics proves insufficient to fully describe certain scenarios of its linear propagation. In particular, we show that the (quantum) vacuum imprints unexpected and surprisingly useful information about the dielectric medium that is not captured by $\epsilon^{(1)}(\boldsymbol{r},\omega)$ of the material and manifests already at the electric field amplitude level.\\
\noindent The scenario of interest is depicted in Fig.~\ref{fig: schematic_raman_process}. A single molecule is deposited on the surface of a high-\textit{Q} microtoroid resonator, which is, in turn, coupled to an optical fiber. This setup has become a standard molecular sensing tool in the last decade~\cite{vahala2003optical,heylman2016optical,subramanian2018label,armani2007label,suebka2024ultra,abhyankar2020scalable, heylman2016optical,knapper2018single}: regardless of the coherence properties of the light~\cite{brumer1989one}, its transmission through the fiber acknowledges the presence of the single molecule via a tiny phase shift induced by the molecular permittivity  $\chi^{(1)}(\omega)$~\cite{MukamelBook,chen2020three,kaplan2013finite,swaim2011detection}.  The main feature that makes toroidal microresonators suitable for this formidable task is the unusually long photon lifetime: the ultrahigh $Q$-factors achievable through advanced microfabrication techniques trap the photon for $Q=10^3-10^{10}$ roundtrips, amplifying the weak perturbations produced by the single molecule~\cite{yoshie2011optical,toropov2024thermo,houghton2024whispering,saavedra2025origin}. \\
\noindent More generally, the extraordinary photon lifetimes achievable in these resonators have found application across various fields like enhancing weak signals in gravitational wave detection~\cite{zhou2018optomechanical}, atmospheric particle tracking~\cite{kuznetsova2004microparticle}, and biosensing~\cite{rho2020label,baaske2014single,ozgur2015label}, among other areas. This longevity should also enable the detection of unlikely processes that nonetheless manifest if given enough time. Of our particular interest are rare events mediated by quantum vacuum, such as that depicted in the energy-level diagram of Fig.~\ref{fig: schematic_raman_process}. As the light with frequency $\omega$ passes through the microresonator, it can act as a pump inducing a Stokes Raman scattering event, modifying the vacuum with a red-shifted field at frequency $\omega_c$ while leaving a vibrational coherence at frequency $\omega-\omega_c$ in the molecule. However, due to the high $Q$ of the microresonator, there is a finite probability that this red-shifted field will be repurposed to act as a second pump, thereby regenerating the vacuum and simultaneously inducing an anti-Stokes Raman scattering event. The output light is at the same frequency $\omega$ as the incoming light, its amplitude is proportional to the amplitude of the latter, both features highlighting the linear optical nature of the phenomenon. However, the novelty is that contrary to the classical optics paradigm, it cannot be understood using the linear susceptibility of the material, $\chi^{(1)}(\omega)$, alone, (Raman scattering events can only be described with higher-order susceptibilities involving more than two transition dipole moments)~\cite{schatz2002quantum,mukamel1977resonance}. The resolution to this apparent conundrum is that two of the dipole interactions are mediated by the vacuum.\\

\noindent Importantly, the phenomenon described above is undetectable outside of these highly controlled photonic environments~\cite{vahala2003optical,herr2014mode,gorodetsky1999optical,fields2000nonlinear}. Owing to the three-dimensional nature of the electromagnetic continuum, it is entropically unlikely that the emitted Stokes Raman field will be reabsorbed by the molecule in free space. This is not an issue in the microtoroid resonator where the discreteness of its mode structure~\cite{vahala2003optical}, as well as the longevity of the trapped photon~\cite{Armani2003ultra,vahala_science,Ostby2009photonic}, ``force" the Stokes field to interact again with the molecule.  This constitutes a novel weak-coupling cavity quantum-electrodynamics (c-QED) phenomenon that is quite distinct from the well-known Purcell effect~\cite{Purcell1946spontaneous, petrak2014purcell,Kleppner1981inhibited,noda2007spontaneous,petrak2014purcell,hummer2016cavity}; as opposed to the latter, it benefits from a low density of photon modes, which is precisely what the three-dimensional confinement of the microtoroid resonator offers. \\
\noindent With this preamble, this work leverages confined geometries and prolonged photon lifetimes to capture rare events, introducing a new paradigm for microresonator applications and paving the way for innovative approaches in chemical and biosensing as well as advancements in Raman technology. 
          
\normalsize

% LINEAR RESPONSE --------------------X----------------X--------------------------------
%--------------------X----------------X--------------------------------
% --------------------X----------------X--------------------------------
% --------------------X----------------X--------------------------------
% --------------------X----------------X--------------------------------
% --------------------X----------------X--------------------------------
% --------------------X----------------X--------------------------------

\section{Linear response of microtoroid coupled to a single molecule}

\begin{figure*}[ht!]
\begin{centering}
\includegraphics[width=0.50\textwidth]{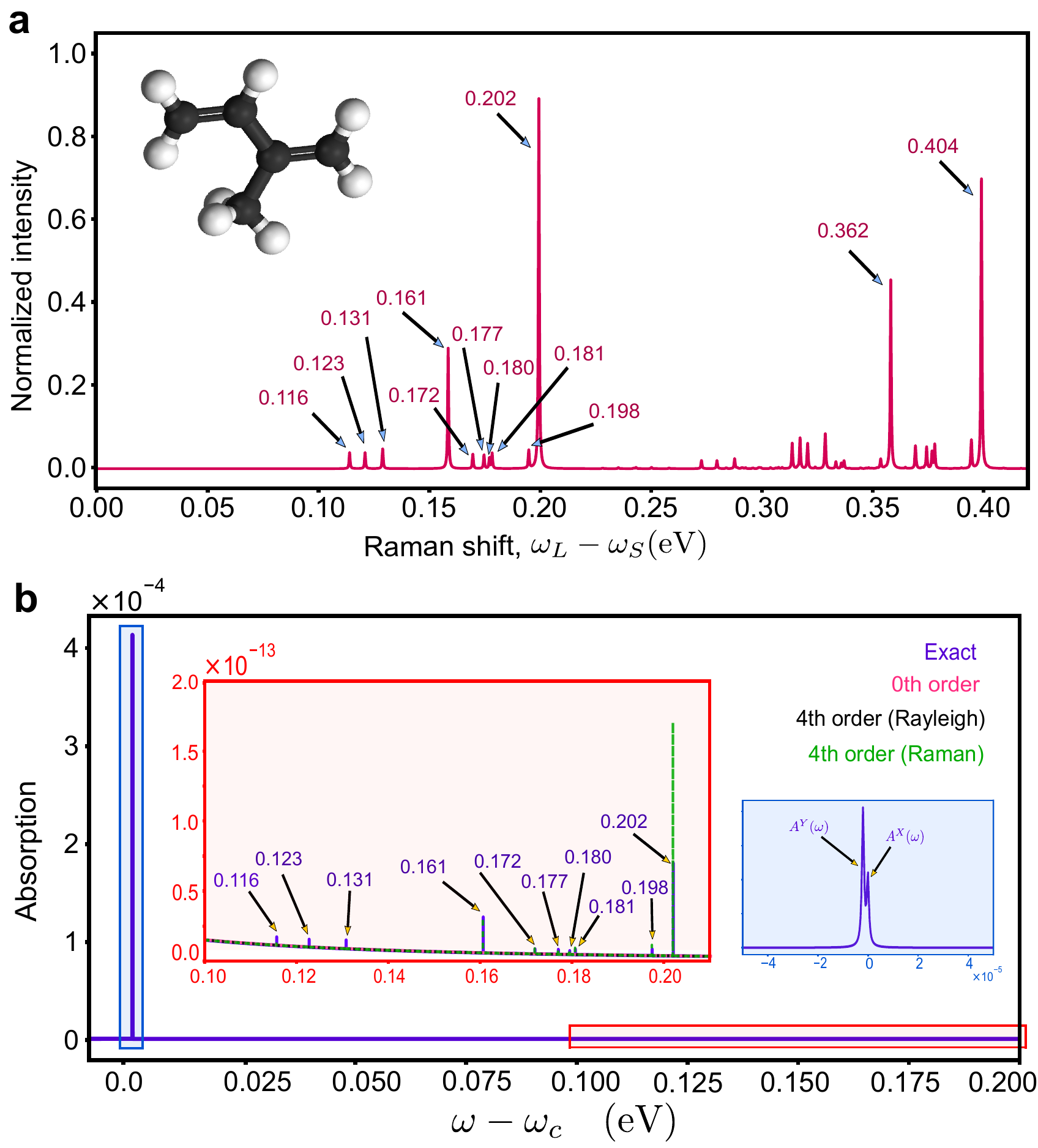}
      
\caption{
\textbf{Cavity vacuum-mediated Raman processes in isoprene encode the information of the bare Raman spectrum.} 
\textbf{a}, Raman spectrum of bare isoprene, where the $y$-axis represents the normalized scattered intensity, and the $x$-axis corresponds to the Raman shift, defined as the frequency difference between incident and scattered light. 
\textbf{b}, Frequency-resolved absorption spectrum of a microresonator coupled to a single isoprene molecule. The exact calculation of the spectra is plotted in violet. The blue inset highlights the lifting of degeneracy in the microresonator modes, leading to symmetric and anti-symmetric combinations. The red inset reveals additional peaks attributed to cavity vacuum-mediated Raman processes near the cavity resonance at $\omega=0$, as depicted in Fig.~\ref{fig: schematic_raman_process}. The $0^{\text{th}}$ order term (magenta) in Eq.~\ref{eqn:pert_4_one_isoprene} reproduces the bare cavity spectrum. The fourth-order term in the Dyson expansion is split into two parts. The first part (black) includes only Rayleigh features, encoding only the linear susceptibility, $\chi^{(1)}(\omega)$, of the molecule and fall in the paradigm of classical linear optics. In contrast, the second part of the fourth-order(green) term unveils new, subtle peaks around the cavity resonance—signatures of quantum vacuum-mediated Stokes-anti Stokes Raman processes that introduce additional absorption channels at frequencies detuned from the cavity resonance by the Raman shifts. These peaks do not exist in the absorption spectra outside the cavity and are signatures of the cavity vacuum-mediated weak coupling effect presented in this work.}
\label{fig:one_isoprene_raman}
\end{centering} 
\end{figure*}

%\noindent \textcolor{blue}{A solid para on the experimental realisability of this phenomena to be written}\\

\noindent High-$Q$ microresonators with three-dimensional confinement, such as microspheres and microtoroids, are emerging as a frontier in sensing, with a growing effort towards operation across a diverse spectral range. When coupled to a molecule, this optical confinement in three dimensions leads to a discrete mode structure instrumental to the quantum-vacuum mediated Raman, while the high-Q nature of the resonator allows enhanced sensitivity to resolve the subtle spectral features in the linear spectra. Here, we establish a theoretical framework that rigorously captures both the optical mode of a microtoroid resonator and the quantum dynamics of the molecule. By employing an input-output formalism, we compute the frequency-resolved transmission spectrum for arbitrary molecular complexity, revealing molecular Raman signatures mediated through quantum-vacuum and emerging at the spectral tails. \\
\noindent The interaction between an optical toroidal microresonator and a single molecule can be effectively described by a reduced model when the frequency range of interest is narrower than the free spectral range (FSR). Under these conditions, the Hamiltonian for the setup is given as~\cite{vahala_science}, 

\begin{eqnarray}
H=\sum_{\alpha=a,b}H_{\text{ph},\alpha}+H_{I}+H_\text{mol}+V,\label{microtoroid_N_molecules_Hamiltonian}
\end{eqnarray}
\\
where, in natural units, $H_{\text{ph},\alpha} = \omega_c \alpha^{\dagger} \alpha$ governs the two degenerate counter-propagating optical modes arising due to the intrinsic symmetry of the resonator; $\alpha = a$ ($b$) denotes the annihilation operator for the clockwise (counterclockwise) mode. Scattering processes induced by structural imperfections in the resonator couple these modes via $H_{I} = \beta (a^\dagger b + b^\dagger a)$, with $\beta$ denoting the coupling strength. Further, these optical modes couple to a molecule via an evanescent interaction, which, under the rotating wave approximation (RWA), is described by  
\begin{eqnarray}  
V = g(a^\dagger + b^\dagger)\sigma + h.c.,  
\end{eqnarray}  
where $ \sigma $ denotes the lowering operator associated with the electronic transition of the molecule. The parameter $ g = \sqrt{\frac{\omega_c}{2\epsilon_0 \mathcal{V}_m}} $ represents the single-molecule light-matter coupling strength, with $ \mathcal{V}_m $ being the effective mode volume and $ \epsilon_0 $ the permittivity of free space. Since $V$ preserves the total excitation number, the molecular Hamiltonian, in the linear response regime, is constrained to the first excitation manifold and is given by,
\begin{align}  
H_\text{mol}=\sum_{n}\omega_{g, n}\left|g,\psi_n\right\rangle \left\langle g,\psi_n\right|+\sum_{m}\omega_{e,m}\left|e,\varphi_m\right\rangle \left\langle e,\varphi_m\right|,  
\end{align}  
where $ |g,\psi_n\rangle $ and $ |e,\varphi_m\rangle $ denote the quantum states of the molecule in the ground and excited vibronic manifolds, with energies $ \omega_{g,n} $ and $ \omega_{e,m} $ respectively. \\
\noindent For high-$Q$ cavities, the reflection, $R(\omega)$, is inherently minimal (supplementary materials section 1), making transmission, $T(\omega)$, the dominant experimentally accessible signal~\cite{Armani2003ultra,Fang2024million}. Conservation of photon flux enables us to express the frequency-resolved transmission in terms of the absorption, $T(\omega) = 1- A(\omega)$~\cite{yuen2024linear}. We will primarily present absorption spectra throughout this work to facilitate direct comparisons with bare molecular spectra.\\
\noindent The absorption spectrum, $A(\omega)$, of the microresonator coupled to a single molecule can be rigorously obtained using input-output theory (supplementary materials section 1,2) in terms of the  photon Green's function~\cite{steck2007quantum,vahala_science}. Since only the symmetric superposition of the optical modes of the resonator interacts with the molecule, it is convenient to define the annihilation operators \( X = \frac{a + b}{\sqrt{2}} \) and \( Y = \frac{a - b}{\sqrt{2}} \), corresponding to the symmetric and antisymmetric modes, respectively. This leads to the linear absorption of the microtoroid–molecule system being decomposable into distinct contributions (symmetric and anti-symmetric), 
\begin{eqnarray}
A(\omega) &=& \sum_{\alpha = X,Y} A^{\alpha}(\omega), 
\end{eqnarray}
with 
\begin{eqnarray}
    A^{\alpha}(\omega) &=& -\kappa_{\text{ex}}\Im\big[D^{\alpha}(\omega)\big] - \frac{\kappa_{\text{ex}}^{2}}{2}\big|D^{\alpha}(\omega)\big|^{2}.
\end{eqnarray}
\\
Here $D^{\alpha}(\omega)=\langle 1_{\text{ph}}^\alpha;g,\psi_0|G(\omega)|1_{\text{ph}}^\alpha ;g,\psi_0\rangle$ is the matrix element of the total Green's function, $G(\omega)=\frac{1}{\omega-H+i0^+}$, corresponding to the quantum state $|1_{\text{ph}}^\alpha ; g,\psi_0\rangle$ with a photon in the $\alpha$ mode of the cavity and the molecule in the global ground state~\cite{cwik2016excitonic,zeb2018exact,yuen2024linear}. Further, the field decay rate for the resonator normal modes is $\kappa = \kappa_\text{i} + \kappa_\text{ex}$, where $\kappa_\text{i}$  represents intrinsic losses and $\kappa_\text{ex}$ describes extrinsic loss due to (adjustable) couplings of the modes to the fiber coupler \cite{vahala_science}. \\
\noindent Since the antisymmetric optical mode remains decoupled from the molecular bath, its absorption component,
\begin{eqnarray} A^{Y}(\omega) = \frac{\kappa_{\text{ex}}\kappa_{i}}{\bigl(\omega - (\omega_{c} - \beta)\bigr)^{2} + \frac{\kappa^2}{4}},
\end{eqnarray}
is dictated solely by the resonator’s intrinsic dissipation, producing a characteristic Lorentzian profile. In contrast, the absorption of the symmetric mode, $A^{X}(\omega)$, encodes modifications in the Lorentzian lineshape arising from molecular interactions, imprinting a spectral fingerprint of the molecule within the response of the resonator~\cite{richter2018microtoroid,Min2006UltrahighQ}.\\
\noindent Contrary to $A^Y(\omega)$, the computation of $A^X(\omega)$ is non-trivial due to the inherent complexity of the molecule~\cite{yuen2023linear,cwik2016excitonic}. In Fig.~\ref{fig:one_isoprene_raman}, we present the numerically computed $A^{X}(\omega)$ using the exact photon Green's function $D^{X}(\omega)$ (supplementary materials section 5). For this calculation, we employed isoprene, a well-established benchmark for studying linear and nonlinear optical responses due to its well-characterized electronic structure and biological relevance, making it an ideal candidate for validating theoretical and experimental methods in spectroscopy and photophysics~\cite{Heller_isoprene,gao2021study,Sahay2013measurements,zhan2021assessment}. Its electronic transition occurs at a characteristic frequency of $\omega_{0-0} = 6.026$ eV with an oscillator strength of \(0.5523\), while its vibrational energy landscape has been parametrized with \(10\) displaced harmonic oscillators \cite{Heller_isoprene,NIST_C78795} (supplementary materials Table 2). Furthermore, this molecule is evanescently coupled to the $m = 27^\text{th}$ mode of the microtoroidal resonator, which is red-detuned by $\Delta  = 0.3$ eV relative to $\omega_{0-0}$.
 Under these conditions, the light-matter coupling strength is determined to be $g = 1.63 \times 10^{-4}$ eV (supplementary materials section 7), based on an approximated mode volume \cite{Ostby2009photonic, cai2020whispering,benson2006micro}. This parameter set corresponds to the weak coupling regime, as the ratio $g/\Delta$ remains much smaller than unity. Intriguingly, an additional set of peaks emerges near the cavity resonance in absorption spectra (Fig. ~\ref{fig:one_isoprene_raman}), which cannot be explained by Purcell effect, CERS, or other known cavity-induced weak coupling effects~\cite{Purcell1946spontaneous, petrak2014purcell,Kleppner1981inhibited,Haroche1992cavity,Jhe1987SuppressionOS,andrew2004energy,hotter2023cavity,Fujii2004CavityQED,Giuliano2024thermal,Noh2020emission,yoshie2004vacuum,noda2007spontaneous}. However, these new peaks can be captured by a Dyson expansion of the photon Green’s function, $D^{\alpha}(\omega)$, in the light-matter coupling $V$.\\
 \noindent The photon Green's function up to the fourth order term in the Dyson series, 
\begin{eqnarray}
  D^{X}(\omega)= D_{(0)}^{X}(\omega)+D_{(2)}^{X}(\omega)g^2+D_{(4)}^{X}(\omega)g^4 + \mathcal{O}(g^6),
\label{eqn:pert_4_one_isoprene}
\end{eqnarray}

\noindent is sufficient to explain the appearance of the new peaks in Fig. \ref{fig:one_isoprene_raman} near the cavity resonance. Higher-order contributions merely rescale the intensity of existing features rather than introducing new ones (see supplementary materials section 4). Since $D^{\alpha}(\omega)$ is a diagonal component of the total Green’s function $G(\omega)$, all odd-order terms in the Dyson expansion vanish, leaving only even-order corrections.~\cite{dittes2000decay,MukamelBook} \\

\noindent In the absence of molecules, the zeroth-order cavity response, $D_{(0)}^{X}(\omega) = 1/(\omega - \omega_c + i \frac{\kappa}{2})$, accounts for the Lorentzian profile of the bare microtoroid resonator~\cite{Armani2003ultra,vahala_science,Ostby2009photonic}. With the introduction of light-matter coupling, the second-order contribution is $ D_{(2)}^{X}(\omega) = -\, g^2 \left[ D_{(0)}^{X}(\omega) \right]^2 \chi^{(1)}(\omega)$; here $\chi^{(1)}(\omega)=- \lim_{\gamma \to 0^+}\sum_{n=0}\frac{|\left\langle \psi_{0}\big|\varphi_{n}\right\rangle |^{2}}{\omega-\omega_{e,n}+i\frac{\gamma}{2}}$  is the linear susceptibility of the molecule representing the cavity mediated Rayleigh-type scattering processes through which energy is dissipated into the molecular bath~\cite{MukamelBook,tannor2007introduction}. Here, $\gamma \to 0^+$ ensures the causality of the Green’s function and can be chosen arbitrarily small in simulations as the molecular complexity is explicitly captured by introducing sufficient vibrational modes~\cite{chin2010impurity,Gin2007hierarchy,cederbaum2005short}. This second-order term {is captured in classical optics and} reveals how molecular interactions perturb the Lorentzian profile of the bare resonator, introducing an additional decay channel driven by the dipole coupling to the molecules. \\

\noindent Finally, the fourth-order term  $D_{(4)}^{X}(\omega) = g^4 \left[D_{(0)}^{X}(\omega)\right]^3 \left[\chi^{(1)}(\omega)\right]^2 + R_\text{vib}(\omega)$  modifies the cavity lineshape in two distinct ways. The first contribution incorporates the {classical optical} effects of higher-order Rayleigh scattering in the spectral response, whereas the second term,
\small
\begin{align}
R_\text{vib}(\omega) = g^4 \left[D_{(0)}^{X}(\omega)\right]^2 \sum_{m=1}^{M_g} \frac{\left\langle \psi_0 \right| G_\text{ex}(\omega) \left|\psi_m \right\rangle \left\langle \psi_m \right|G_\text{ex}(\omega) \left|\psi_0 \right\rangle}{\omega - \bigl(\omega_c + \omega_{g,m}\bigr) + i \,\tfrac{(\kappa + \gamma_{\text{vib}})}{2}},
\label{eqn:one_isoprene_raman_before_approx}
\end{align}
\normalsize
{is beyond classical optics and }describes interactions involving the cavity photon and ground state molecular vibrations of frequency $\omega_{g,m}$ with an associated decay rate $\gamma_\text{vib}$; here $G_\text{ex}(\omega)=\sum_m\frac{|\varphi_m\rangle\langle\varphi_m|}
{(\omega-\omega_{e,m}+i\frac{\gamma}{2})}$ is the excited state Green's function of the molecule. These processes generate additional spectral features near the cavity resonance, $\omega_c$, as shown in Fig.~\ref{fig:one_isoprene_raman}, that are not captured by lower-order treatments. Importantly, Eq. \ref{eqn:one_isoprene_raman_before_approx} approximated near the cavity resonance, $R_\text{vib}(\omega)\approx g^{4}N\Big[d_{N,0}(\omega)\Big]^{2}S_{\text{Raman}}(\omega_{L}=\omega,\omega_{S}=\omega_{c})$ captures the same information as the Stokes Raman spectra in Fig. 2a where, 
\begin{equation}
    S_{\text{Raman}}(\omega_L, \omega_S) = 2\pi \sum_{m=1}^\infty \left| \langle \psi_0 | G_\text{ex}(\omega) | \psi_m \rangle \right|^2 \delta(\omega_S - \omega_L + \omega_{g,m})
\end{equation} 
denotes the Kramers-Dirac-Heisenberg (KDH) Stokes Raman cross-section~\cite{MukamelBook,schatz2002quantum,long2002raman} with a laser of frequency $\omega_L$  undergoing inelastic scattering to leave behind a vibrationally hot molecule and a red-shifted Stokes field at  $\omega_S = \omega_L - \omega_{g,m}$ (see Fig.~\ref{fig: schematic_raman_process}).
To highlight the underlying mechanism,  Eq.~\ref{eqn:one_isoprene_raman_before_approx} can be reorganised into (supplementary materials section 4)
%\begin{widetext}
 \begin{eqnarray}
     R_\text{vib}(\omega)\approx\frac{g^{4}}{2\pi}\Big[D_{(0)}^{X}(\omega)\Big]^{2}\sqrt{S_{\text{Raman}}(\omega_{L}=\omega,\omega_{S}=\omega_{c})}\sqrt{S_{\text{Raman}}^{\text{C}}(\omega_{L}=\omega_{c},\omega_{AS}=\omega)},
     \label{eqn:one_isoprene_raman_after_approx}
 \end{eqnarray}
%\end{widetext}

\noindent where the conditional KDH anti-Stokes Raman cross-section,  
\begin{equation}
    S^C_{\text{Raman}}(\omega_L, \omega_{AS}) = 2\pi \sum_{m=1}^\infty \left| \langle \psi_m | G_\text{ex}(\omega) | \psi_0 \rangle \right|^2 \delta(\omega_{AS} - \omega_L - \omega_{g,m})
\end{equation}  
describes the inverse scenario where an incident photon of frequency $\omega_L$ inelastically scatters by extracting vibrational energy from the molecule leaving behind a vibrationally cold molecule at the global ground state, while the scattered photon emerges blue-shifted at $\omega_{AS} = \omega_L + \omega_{g,m}$ as illustrated in Fig. ~\ref{fig: schematic_raman_process}\\

\begin{figure*}[ht!]
     
        \includegraphics[width=1 \textwidth]{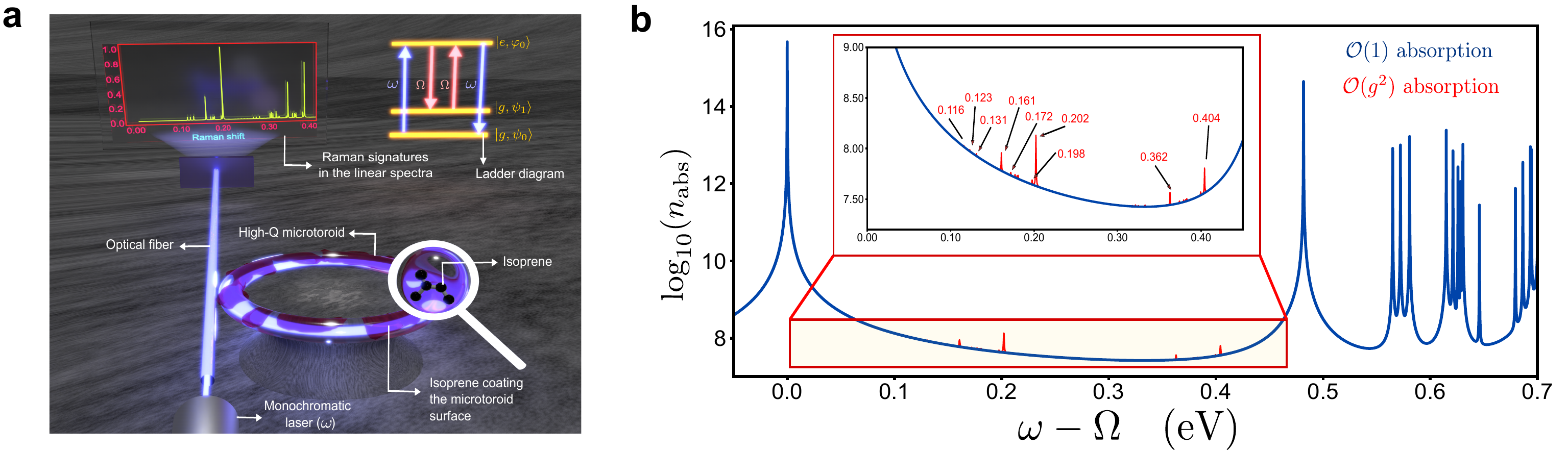}
        
\caption{\textbf{Proposed experimental protocol for enhanced Raman peak detection in the linear transmission spectra. }
\textbf{a}, Schematic of the experimental protocol for Raman peak enhancement, where an ensemble of isoprene molecules are evanescently coupled to a microtoroid resonator (highlighted by the magnifying glass). An optical fiber delivers the laser, which couples evanescently to the resonator, thereby imprinting the molecular Raman signatures onto the absorption spectrum in the linear regime. 
% (should we put Transmission?)   
Corresponding ladder diagram illustrating the energy-level transitions, \(\Omega\) assuming the role that $\omega_c$ took in the single-molecule case.
\textbf{b}, The main plot presents the logarithm (base 10) of the number of absorbed photons, \( \log_{10} n_{\text{abs}} \), as a function of \( \omega - \omega_{\text{LP}} \), where \( \omega_{\text{LP}} \) corresponds to the lower peak of the absorption spectrum. The system parameters are set as \( \Delta = 0 \), \( g = 1.63 \times 10^{-4} \) eV, \( N = 3\times10^6 \), \( \kappa_{\text{ex}} =\gamma = 3 \times 10^{-5} \) eV, and \( \gamma_{\text{vib}} = 5 \times 10^{-4} \) eV (considering $\omega_c=\omega_{0-0}$, this corresponds to $Q=2\times10^5$~\cite{perin2022high,Min2006UltrahighQ,Liu18ultraviolet,richter2018microtoroid}). The inset highlights the second-order correction to the absorption, revealing distinct Raman signatures. Both the fundamental Raman band and its higher-order overtones emerge, suggesting that vacuum-mediated effects play a role in the observed vibrational features. Notably, the photon absorption remains well above the noise equivalent power (NEP, \( 10^5 \)), with \( n_{\text{abs}} > 10^7 \). 
}
\label{fig:protocol_N_isoprene}
        \end{figure*}
        
\noindent Eq.~\ref{eqn:one_isoprene_raman_after_approx} represents the underlying vacuum-mediated Stokes and conditional anti-Stokes processes, as illustrated in Fig.~\ref{fig: schematic_raman_process}, using the Double-Sided Feynman Diagram (DSFD). The incoming laser field at frequency $\omega_{L} = \omega$ drives a Stokes scattering process, leaving behind a scattered field at the cavity resonance, $\omega_{S} = \omega_c$. This scattered field, in turn, serves as the incident field for a constrained anti-Stokes process at $\omega_{L} = \omega_c$, ultimately yielding a final scattered field at the original laser frequency, $\omega_{AS} = \omega$. This new vacuum-mediated phenomenon introduces additional absorption channels at frequencies detuned from the cavity resonance by the Raman shifts, thus harvesting Raman signatures already in the linear response of the microtoroidal resonator.\\
\noindent The presence of multiple noise sources, including photon shot noise, which scales as $\propto \sqrt{N_\text{ph}}$~\cite{RPPhotonics2025ShotNoise}, and the intrinsic dark current of photodetectors, typically on the order of $10^5$ photons for silicon photodiodes at room temperature~\cite{Thorlabs2023DarkCurrent}, poses significant challenges for the experimental observation of these subtle spectral features. Given that their strength is on the order of $10^{-13}N_{\text{ph}}$, achieving a signal above the photon shot-noise would require averaging over $10^{26}$ photons. Such an intense photon flux risks inducing spurious effects due to photothermal effects and nonlinear optical processes~\cite{Hamamatsu2021SiPD, IEEE2016high}, and also exceeds the saturation thresholds of conventional detectors~\cite{HST2020,Thorlabs2020photodiode,Hamamatsu2022,Zhao2024}.  In the following section, we introduce a detection scheme designed to enhance the signal-to-noise ratio by leveraging collective interactions for a molecular ensemble, thereby proportionally amplifying the detected signal and mitigating the limitations imposed by detector and shot noise.

 \begin{figure*}[ht!]
 \begin{centering}
 \includegraphics[width=0.75 \textwidth]{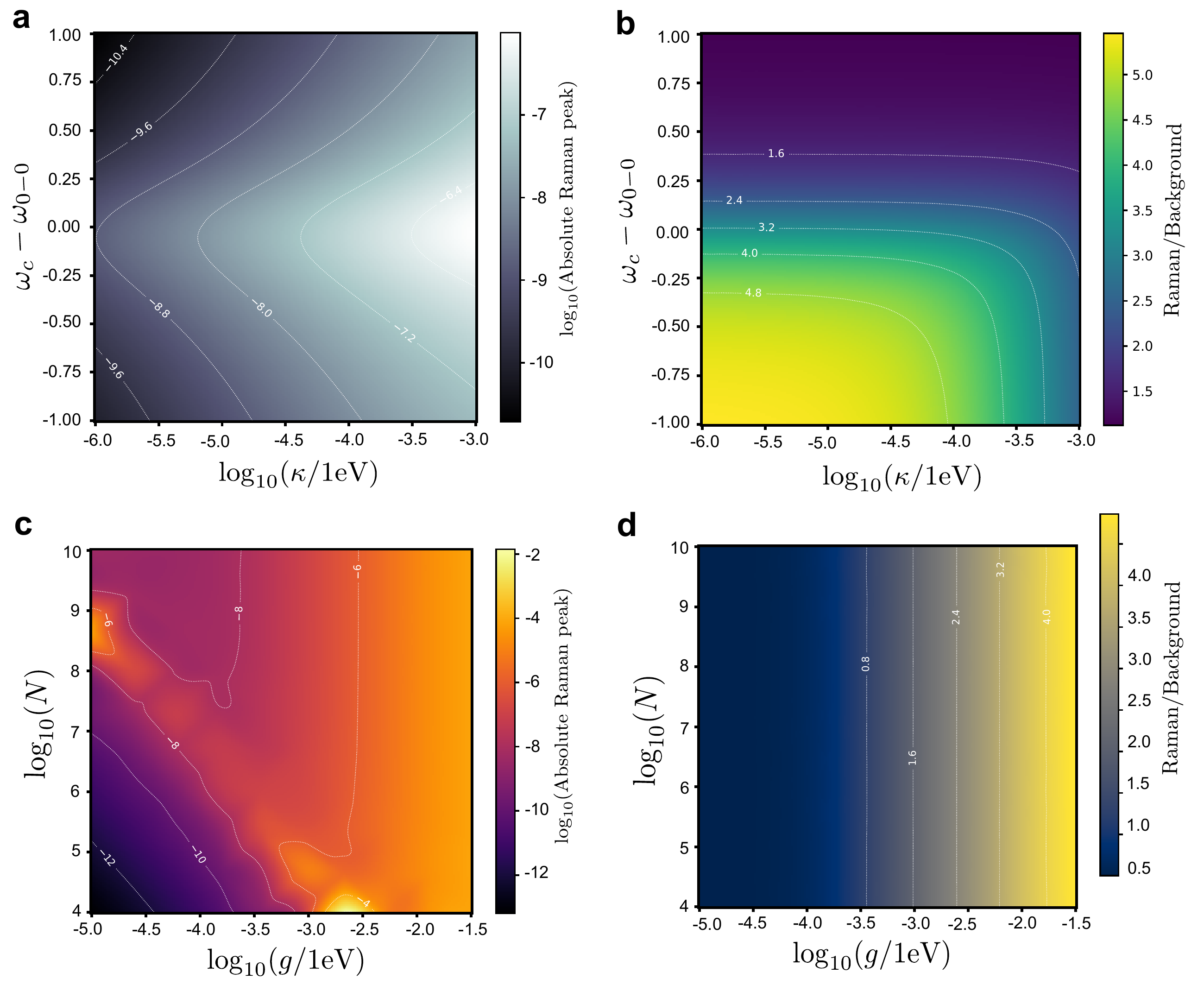}
\caption{\textbf{Dependence of absolute Raman peak height and its ratio to the Rayleigh background on key experimental parameters}.   Contour plots of
\textbf{a}, the absolute Raman peak height as a function of $\Delta=\omega_c-\omega_{0-0}$ and $\log_{10}(\kappa/\text{1 eV})$,
\textbf{b}, the ratio of Raman peak height to the Rayleigh background over the same parameter space with \( g = 1.63 \times 10^{-4} \) eV, \( N = 3\times10^6 \), \( \gamma = 3 \times 10^{-5} \) eV, and \( \gamma_{\text{vib}} = 5 \times 10^{-4} \) eV for this simulation. 
\textbf{c}, the absolute Raman peak height as a function of $\log_{10}(g/\text{1 eV})$ and $\log_{10}(N)$, and
\textbf{d}, the ratio of Raman peak height to the Rayleigh background over the same parameter space with \( \kappa_{\text{ex}} = \gamma = 3 \times 10^{-5} \) eV and \( \gamma_{\text{vib}} = 5 \times 10^{-4} \) eV for this simulation,   
These simulations focus on the dominant Raman feature from the vibrational mode with frequency ($\omega_v=0.202$ eV) and Huang-Rhys factor ($S=1.57$). Figs. \textbf{a.} and \textbf{b.} indicate that $Q$ factors in the range of $2\times10^4$ to $2\times10^6$
are optimal for maximizing both the signal strength and the Raman/background ratio.}
\end{centering}
\label{fig:scaling}
        \end{figure*}
       
\section{Detection scheme}

\noindent Enhancements of absorption cross-sections can be directly achieved by increasing the number of absorbers~\cite{beerlambert2023}. In our system, this corresponds to increasing the number of absorbers evanescently coupled to the microtoroidal resonator. The upper limit for the surface density of such absorbers is \( \rho_{\text{max}} = 5 \times 10^{14} \, \text{cm}^{-2} \) \cite{zhuravlev1987concentration}, which, given the estimated inner and outer diameters of the resonator (supplementary materials section 8), translates to a maximum of \(10^7 \) molecules that can be evanescently coupled. However, an exact computation of the photon Green’s function for a resonator interacting with millions of molecules is an intractable problem. Consequently, a Dyson expansion approach provides the only feasible (and well-justified) framework for evaluating the photon Green’s function in this regime~\cite{cwik2016excitonic, zeb2018exact,yuen2024linear}.\\
\noindent For a microtoroid resonator resonantly coupled to $3\times10^6$ isoprene molecules, a Dyson expansion of the photon Green's function in the light-matter interaction $V$ reveals the existence of two dynamical timescales: a timescale corresponding to the collective coupling $g\sqrt{N}$ that drives the Rayleigh scattering processes and a single-molecule timescale, $2\pi / g$, corresponding to the sought-after Raman processes. To address this inherently multiscale problem, we incorporate the collective effects exactly via a self-consistent Dyson equation (supplementary materials section 6) while treating \( g \)-mediated single-molecule interactions perturbatively. \\
\noindent As a result, the  zeroth-order photon Green’s function is modified by the fast Rayleigh transitions,  
\[
D_{(0)}^{X(N)}(\omega) = \frac{1}{\omega - \omega_c + i \frac{\kappa}{2} + N \chi^{(1)}(\omega)}
\]
\noindent naturally embeds the linear molecular susceptibility of the ensemble, capturing all orders of the collective Rayleigh processes (which are all the effects in classical linear optics). Beyond this, the second-order term in the expansion,  

%\begin{widetext}
\begin{eqnarray}
D_{(2)}^{X(N)}(\omega) \propto g^2\Big(g\sqrt{N}\Big)^2\Big[D_{(0)}^{X(N)}(\omega)\Big]^{2}\sqrt{S_{\text{Raman}}(\omega_{L}=\omega,\omega_{S}=\Omega)}\sqrt{S_{\text{Raman}}^{C}(\omega_{L}=\Omega,\omega_{AS}=\omega)}
\label{eqn:N_isoprene_raman_after_approx}
\end{eqnarray}
%\end{widetext}

\noindent evaluated around the lower absorption peak, \( \Omega \) (modified by collective interactions), can be expressed in terms of Stokes and constrained anti-Stokes Raman cross-sections, akin to Eq.~\ref{eqn:one_isoprene_raman_after_approx}. However, two key distinctions emerge. First, in contrast to the single-molecule case, where the Stokes-scattered field, serving as the pump for the ensuing anti-Stokes transition, is generated at $\omega_c$, the corresponding field in this regime manifests at frequency $\Omega$. (see DSFD in Fig.~\ref{fig:protocol_N_isoprene} for visualization). This results in Raman signatures emerging along the spectral tails of the Lorentzian profile centered at \( \Omega \) instead of $\omega_c$ as illustrated in Fig~\ref{fig:protocol_N_isoprene}. More importantly, the resulting Raman peaks from the exact expression for $D_{(2)}^{X(N)}(\omega)$ reveal an enhancement of the signal strength, effectively overcoming noise contributions in the spectrum with a sufficient number of molecules.\\

\noindent The Raman signal now reaches around $10^8$ photons, exceeding the intrinsic dark current of photodetectors~\cite{Thorlabs2023DarkCurrent} while remaining below saturation~\cite{HST2020}, as only $10^{16}$ photons need to be integrated for shot-noise-limited measurements. However, rapid detection by focusing $\sim10^{16}$ photons/s into a micrometer-scale fiber demands precise alignment and may introduce optical damage or nonlinear photothermal effects~\cite{Hamamatsu2021SiPD}. Alternatively, lowering flux and extending integration might reduce high-intensity issues but heighten susceptibility to low-frequency drifts~\cite{lowfreq2014measurement} and $1/f$ noise~\cite{Kiely2019analog}. Nonetheless, $\mathcal{O}(10^{-8})$ absorption detection is well within reach using active stabilization~\cite{descloux20213d}, lock-in amplification~\cite{Zurich2025lock}, balanced detection~\cite{robinson2012balanced}, and modulation-demodulation techniques~\cite{HerdaRadio_Demodulation} while heterodyne techniques~\cite{Jin2021f,Feng2019optimized} and short repetitive scans shift signals~\cite{li2018uv,Hu_2021f} beyond 1/$f$ noise. Leveraging these strategies enables the precise extraction of these vacuum-mediated effects from technical noise in state-of-the-art and emerging photonic architectures~\cite{Hao2024,suebka2024ultra,Koppenhofer2023}.
\noindent Moreover, given the multitude of parameters in Eq.~\ref{eqn:N_isoprene_raman_after_approx}, our detection scheme enables precise tuning to further enhance Raman peak visibility. In Fig.~\ref{fig:scaling}, we illustrate how the ratio of the Raman signal to its background—arising from Rayleigh Lorentzian tails—varies with the cavity detuning, $\Delta$, and its linewidth, $\kappa$. As $\kappa$ decreases, both the Rayleigh and Raman peaks become spectrally sharper, thereby improving this ratio. However, the trade-off is that the absolute Raman peak heights decrease in Fig.~\ref{fig:scaling} as they are both filtered by and superimposed on the Rayleigh background. A similar dependence emerges upon tuning $\Delta$ away from resonance since detuning modifies the linewidth in a manner analogous to changes in $\kappa$ (supplementary materials section 6). Together Figs. \textbf{a} and \textbf{b} demonstrate that $Q$ factors between $2\times10^4$ and $2\times10^6$ provide the most favorable conditions for enhancing Raman signal strength while improving the Raman/background ratio.
 \\
\noindent Additionally, in Fig.~\ref{fig:scaling}, we further consolidate our previous understanding into the roles of $g$ or $N$ in improving Raman peak visibility. As anticipated, increasing $N$ raises the absolute Raman intensity since both the Rayleigh and the Raman contributions are enhanced. In contrast, the signal-to-background ratio only improves with increasing $g$, consistent with the $g^4$ scaling of the Raman response in Eq.~\ref{eqn:N_isoprene_raman_after_approx}.\\
\noindent Notably, our model does not explicitly account for the inhomogeneous broadening of the molecular ensemble, as the Raman features near $\Omega$ lie well outside the Gaussian envelope associated with the inhomogeneous component of the bare molecular absorption. This phenomenon—where a sufficiently detuned probe predominantly senses the homogeneous component due to the rapid decay of Gaussian tails—is widely documented in the literature \cite{houdre1996vacuum, schwennicke2024extracting}.

\section{Summary and outlook}

\noindent Our work explores a new frontier in photonic sensing and spectroscopy, where rare quantum vacuum fluctuations imprint Raman signatures in linear optical response, rendering the signal independent of light's coherence properties~\cite{brumer1989one,TannorRice} and hence pave the way for cost-effective and simplified Raman technology. %that operates with monochromatic incoherent sources like LEDs~\cite{brumer1989one,MukamelBook,tannor2007introduction,TannorRice,BrumerShapiro}.
It is important to note that, owing to the off-resonant Raman scattering character of the phenomenon of interest, non-Condon (Herzberg-Teller) effects can also yield a substantial boost in the desired signal~\cite{heller2016theory,osawa1994charge}. The precise dependence of our signal on these contributions will be explored in future studies. Notably, unlike conventional nonlinear Raman techniques, our protocol inherently circumvents the common issue of fluorescence background by strictly prohibiting electronic population generation~\cite{wei2015fluorescence,MukamelBook}. Consequently, the inherently weak Raman signal remains unobscured and free from the masking effects of fluorescence. The appearance of the Raman signal in the linear spectra, however, introduces an inherent competition with classical optical absorption, a challenge that can be effectively mitigated by employing off-resonant microresonators. The detuning from the optical absorption also prevents deleterious photoinduced processes in the analyte from the laser excitation. %Additionally, our approach removes the need for Rayleigh filtering, facilitating the detection of low-frequency Raman shifts without specialized filters—a key challenge in terahertz (THz) Raman spectroscopy~\cite{futamata1996dielectric,Carriere2013TerahertzRaman,Coherent2022IntroTHzRaman}.
The parameter space explored in this work (e.g., free spectral range (FSR), mode volume, etc.) aligns with emerging microresonator technologies, suggesting the full utilization of the quantum vacuum effect with next-generation photonics~\cite{He2023ultra,Liu18ultraviolet,perin2022high}. Nevertheless, state-of-the-art microresonators in the mid-IR to UV regimes, with FSRs in the range of \(5 \text{–} 10\) meV and ultrahigh \(Q\)-factors \(10^{8} - 10^{10}\), offer an ideal platform for THz Raman spectroscopy~\cite{randy_nat_photonics}, significantly simplifying experimental protocols by reducing requirements for fluorescence suppression and Rayleigh filtering. The decrease in signal strength due to the increased mode volume in these mid-IR resonators can be compensated by the enhancement from the increased surface area, allowing for the deposition of a larger ensemble of molecules (see Eq.~\ref{eqn:N_isoprene_raman_after_approx} and supplementary materials section 6). Beyond increasing the molecular ensemble, this platform offers a compelling opportunity to integrate emerging quantum light interferometry techniques, potentially amplifying both signal strength and sensitivity, which is especially relevant for detecting weak signals in low-frequency regimes~\cite{tse2019quantum,giovannetti2011advances,Pizzi2023light}.\\

\noindent From a broader perspective, harnessing the long photon lifetimes in high-\(Q\) microresonators to probe the cavity QED rare events reported here is reminiscent of advances across diverse scientific fields—ranging from neutrino detection (Super-Kamiokande, IceCube)~\cite{murase2015testing,Fukuda1998evidence} and gravitational wave observation (LIGO, Virgo)~\cite{thorne1995gravitationalwaves} to rare decay experiments~\cite{higashino2021weak,batley2002arxiv} and investigations of rare biochemical reactions~\cite{daigle2011automated}—where extended observation times enhance sensitivity and facilitate the capture of elusive, low-probability events. 
In the ensemble protocol, in addition to the quantum vacuum–mediated third-order Raman susceptibility appearing as the leading-order quantum correction to the classical optics, additional features emerge from the higher-order quantum corrections that are not captured by any third-order nonlinear susceptibility~\cite{MukamelBook}(supplementary materials section 6). These electrodynamical events are even rarer and demand emerging ultrahigh-$Q$ microresonators to be observed~\cite{suebka2024ultra,knapper2018phase,randy_nat_photonics}. Additionally, beyond the rotating wave approximation, the weak counter-rotating terms may enable access to higher manifolds and possibly hyper-Raman scattering in the linear spectra~\cite{gia2024detecting,ultra2019forn}. \\
\noindent Looking forward, this work proposes the ambitious goal of leveraging rare quantum-vacuum-mediated events in high-$Q$ resonators coupled to dielectric media, uncovering valuable dynamical information, and pushing the frontiers of advanced sensing and spectroscopic technology.

\section{Acknowledgements}
This work was supported as part of the Center for Molecular Quantum Transduction (CMQT), an Energy Frontier Research Center funded by the U.S. Department of Energy, Office of Science, Basic Energy Sciences under Award No. DESC0021314. A.K. and S.R.C. thank Yong Rui Poh for the Python plotting routine used in Fig. 4. All the authors thank Randall Goldsmith for feedback about the manuscript.

\bibliography{references}
\bibliographystyle{sciencemag.bst}

%---------------------X------------------------X------------------------
% SUPPLEMENTARY MATERIALS
\end{document}

% Insert the title and author list
\maketitle

\section{Input-output theory of a microtoroid resonator coupled to $N$ molecules} \label{sec:input-output_theory}
% \textcolor{blue}{put isoprene modes}

\begin{figure}[htb]
    \centering
    \includegraphics[width=0.75\linewidth]{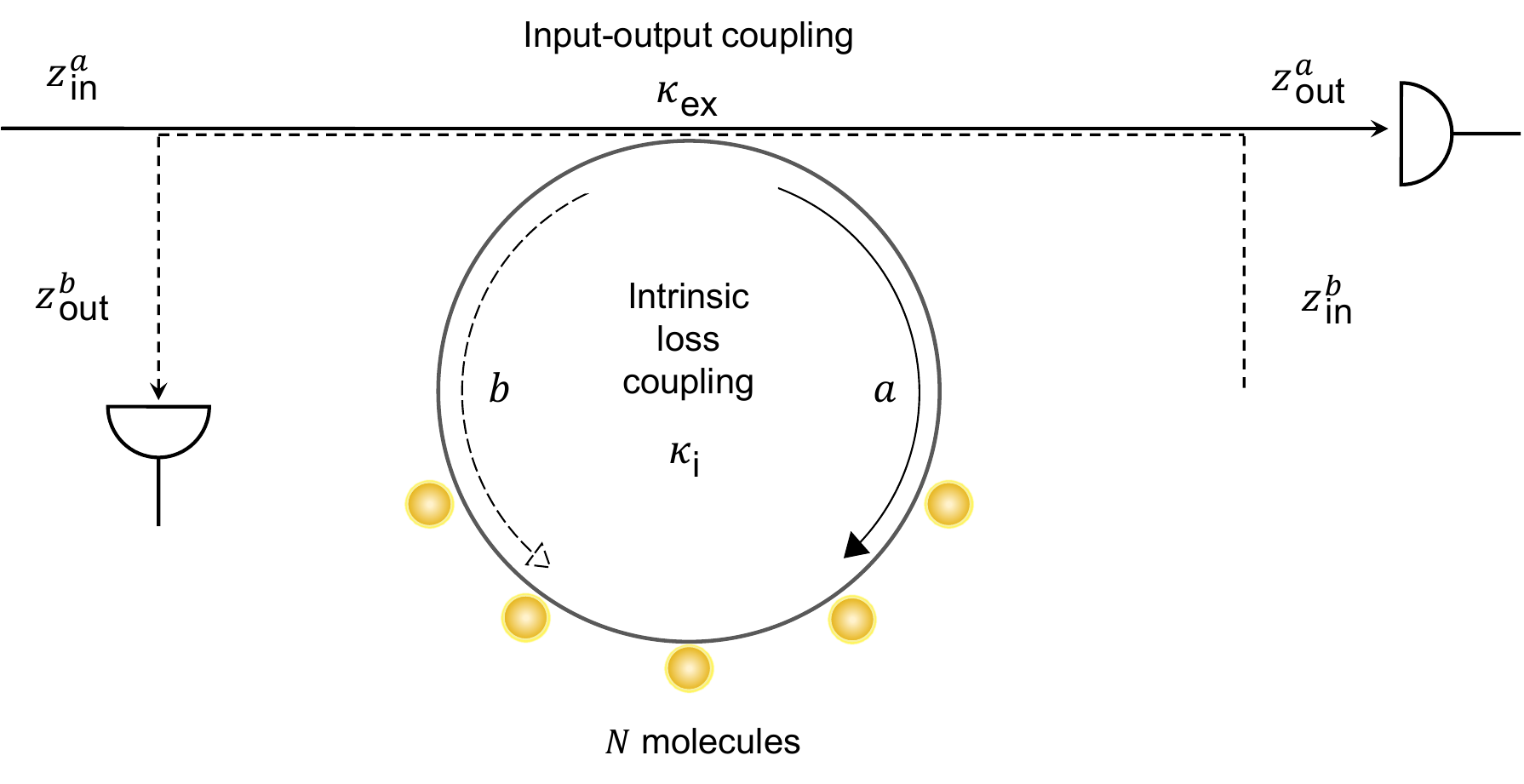}
    \caption{Schematic of microtoroid and fiber coupler}
    \label{fig:enter-label}
\end{figure}

\noindent We derive formulas for linear spectroscopy of microtoroid cavity coupled to $N$ molecules using input-output (IO) theory~\cite{vahala_science, Gardiner1985input,ciuti2006input,steck2007quantum, li2017probing}. We couple this molecular cavity system (m-c-s) with $z^a$ and $z^b$ radiative continua via the photon mode,
\begin{equation}
    H_\text{total} = H + \sum_{\alpha = a,b} H^{z^\alpha},
    \label{eqn:main_hamiltonain}
\end{equation}
\noindent where,
\begin{equation}
    H^{z^{\alpha}}=\hbar\int_{0}^{\infty}d\omega z^{\alpha\dagger}(\omega)z^{\alpha}(\omega)\omega +\hbar\sqrt{\frac{\kappa_{\text{ex}}}{2\pi}}\int_{0}^{\infty}d\omega \alpha z^{\alpha\dagger}(\omega)+ \text{h.c.}
\end{equation}

\noindent for $\alpha = a, b$ only feature Rotating Wave Approximation (RWA) terms and
\begin{subequations}
   
\begin{align}
    [z^{\alpha}(\omega),z^{\alpha'}(\omega')]&=[z^{\alpha\dagger}(\omega),z^{\alpha'\dagger}(\omega')]=0,\\
    [z^{\alpha}(\omega),z^{\alpha'\dagger}(\omega')]&=\delta_{\alpha\alpha'}\delta(\omega-\omega').
\end{align}

\end{subequations}

\noindent For convenience, we will now derive some results in the Heisenberg picture (corresponding to evolution with respect to $H_\text{total}$), with the corresponding operators labeled by the subscript H, e.g., $X_\text{H}(t) = e^{-i H_\text{total}(t-t_\text{in})/\hbar} X e^{i H_\text{total}(t-t_\text{in})/\hbar}$. Schrödinger picture operators will continue to be indicated without an explicit subscript. The Equation of Motion (EoM) for the cavity photon corresponding to the symmetric mode is 
\small
\begin{align}
\nonumber \frac{\partial X_\text{H}(t)}{\partial t}	&=-\frac{i}{\hbar}\Big[X_\text{H}(t),H\Big] \\
	&=-i\big(\omega_{\text{ph}}+\beta\big)X_\text{H}(t)-\frac{\kappa_{i}}{2}X_\text{H}(t)-i\sqrt{2}\sum_{i=1}^{N}g\left|g_{i}\right\rangle \left\langle e_{i}\right|-\frac{i}{\sqrt{2}}\sqrt{\frac{\kappa_{\text{ex}}}{2\pi}}\int_{0}^{\infty}d\omega\big[z_\text{H}^{a}(\omega)(t)+z_\text{H}^{b}(\omega)(t)\big]
\label{eqn:eom_a_h}
\end{align} 
\normalsize
\noindent and the EoM for the $Y_\text{H}(t)$ is
\begin{align}
\nonumber \frac{\partial Y_\text{H}(t)}{\partial t}	&=-\frac{i}{\hbar}\Big[Y_\text{H}(t),H\Big]\\
	&=-i\big(\omega_{\text{ph}}-\beta\big)Y_\text{H}(t)-\frac{\kappa_{i}}{2}Y_\text{H}(t)-i\frac{1}{\sqrt{2}}\sqrt{\frac{\kappa_{\text{ex}}}{2\pi}}\int_{0}^{\infty}d\omega\big[z_\text{H}^{a}(\omega)(t)-z_\text{H}^{b}(\omega)(t)\big].
\label{eqn:eom_b_h}
\end{align}

\noindent Similarly, the corresponding EoM for the bath mode $z^a _\text{H} (\omega) (t)$ is 

\begin{align}
\nonumber \frac{\partial z_\text{H}^{a}(\omega)(t)}{\partial t}	&=-i\omega z_\text{H}^{a}(\omega)(t)-i\sqrt{\frac{\kappa_{\text{ex}}}{2\pi}}\frac{\big[X_\text{H}(t)+Y_\text{H}(t)\big]}{\sqrt{2}}\\
\implies\frac{\partial\big[z_\text{H}^{a}(\omega)(t)e^{i\omega t}\big]}{\partial t}	&=-i\sqrt{\frac{\kappa_{\text{ex}}}{2\pi}}\frac{\big[X_\text{H}(t)+Y_\text{H}(t)\big]}{\sqrt{2}}e^{i\omega t}.
\label{eqn:z_a_eom}
\end{align}

\noindent Defining $t_\text{in} < t$ and $t_\text{out} > t$, we can integrate Eq.~\ref{eqn:z_a_eom} to obtain
\small
\begin{align}
z_\text{H}^{a}(\omega)(t)&=z_\text{H}^{a}(\omega)(t_{\text{in}})e^{-i\omega(t-t_{\text{in}})}-\frac{i}{\sqrt{2}}\sqrt{\frac{\kappa_{\text{ex}}}{2\pi}}\int_{t_{\text{in}}}^{t}dt'X_\text{H}(t')e^{-i\omega(t-t')}-\frac{i}{\sqrt{2}}\sqrt{\frac{\kappa_{\text{ex}}}{2\pi}}\int_{t_{\text{in}}}^{t}dt'Y_\text{H}(t')e^{-i\omega(t-t')},
\label{eqn:input_z_a}
\end{align}
\noindent and
\begin{align}
z_\text{H}^{a}(\omega)(t)&=z_\text{H}^{a}(\omega)(t_{\text{out}})e^{-i\omega(t-t_{\text{out}})}+\frac{i}{\sqrt{2}}\sqrt{\frac{\kappa_{\text{ex}}}{2\pi}}\int_{t}^{t_{\text{out}}}dt'X_\text{H}(t')e^{-i\omega(t-t')}+\frac{i}{\sqrt{2}}\sqrt{\frac{\kappa_{\text{ex}}}{2\pi}}\int_{t}^{t_{\text{out}}}dt'Y_\text{H}(t')e^{-i\omega(t-t')}.
\end{align}
\normalsize
\noindent Similarly, we can obtain expressions for $z^b_\text{H} (\omega)(t)$:
\small
\begin{align}
z_\text{H}^{b}(\omega)(t)&=z_\text{H}^{b}(\omega)(t_{\text{in}})e^{-i\omega(t-t_{\text{in}})}-\frac{i}{\sqrt{2}}\sqrt{\frac{\kappa_{\text{ex}}}{2\pi}}\int_{t_{\text{in}}}^{t}dt'X_\text{H}(t')e^{-i\omega(t-t')}-\frac{i}{\sqrt{2}}\sqrt{\frac{\kappa_{\text{ex}}}{2\pi}}\int_{t_{\text{in}}}^{t}dt'Y_\text{H}(t')e^{-i\omega(t-t')}, 
\label{eqn:input_z_b}
\end{align}
\normalsize

\small
\begin{align}
z_\text{H}^{b}(\omega)(t)&=z_\text{H}^{b}(\omega)(t_{\text{out}})e^{-i\omega(t-t_{\text{out}})}+\frac{i}{\sqrt{2}}\sqrt{\frac{\kappa_{\text{ex}}}{2\pi}}\int_{t}^{t_{\text{out}}}dt'X_\text{H}(t')e^{-i\omega(t-t')}+\frac{i}{\sqrt{2}}\sqrt{\frac{\kappa_{\text{ex}}}{2\pi}}\int_{t}^{t_{\text{out}}}dt'Y_\text{H}(t')e^{-i\omega(t-t')}.
\end{align}
\normalsize
\noindent Let us feed Eq.~\ref{eqn:input_z_a} \& \ref{eqn:input_z_b} into Eq. \ref{eqn:eom_a_h} \& \ref{eqn:eom_b_h} and by approximating $\int_{0}^{\infty}d\omega'\approx\int_{-\infty}^{\infty}d\omega'$, we get,
\begin{subequations} \label{eqn:four_A_B_input_output_relation_source}
 \begin{align}
\frac{\partial X_\text{H}(t)}{\partial t}&=-i\big(\omega_{\text{ph}}+\beta\big)X_\text{H}(t)-\frac{\kappa}{2}X_\text{H}(t)+\sqrt{2}\sum_{i=1}^{N}g\left|g_{i}\right\rangle \left\langle e_{i}\right|-\sqrt{\kappa_{\text{ex}}}\big[z_{\text{in,H}}^{a}(t)+z_{\text{in,H}}^{b}(t)\big],\\
\frac{\partial X_{\text{H}}(t)}{\partial t}&=-i\big(\omega_{\text{ph}}+\beta\big)X_{\text{H}}(t)-\frac{\kappa_{\text{i}}}{2}X_{\text{H}}(t)+\sqrt{2}\sum_{i=1}^{N}g\left|g_{i}\right\rangle \left\langle e_{i}\right|+\frac{\kappa_{\text{ex}}}{2}X_{\text{H}}(t)-\sqrt{\kappa_{\text{ex}}}\big[z_{\text{out,H}}^{a}(t)+z_{\text{out,H}}^{b}(t)\big],
\\
\frac{\partial Y_\text{H}(t)}{\partial t}&=-i\big(\omega_{\text{ph}}-\beta\big)Y_\text{H}(t)-\frac{\kappa}{2}Y_\text{H}(t)-\sqrt{\kappa_{\text{ex}}}\big[z_{\text{in,H}}^{a}(t)-z_{\text{in,H}}^{b}(t)\big],\\
\frac{\partial Y_{\text{H}}(t)}{\partial t}&=-i\big(\omega_{\text{ph}}-\beta\big)Y_{\text{H}}(t)-\frac{\kappa_{\text{i}}}{2}Y_{\text{H}}(t)+\frac{\kappa_{\text{ex}}}{2}Y_{\text{H}}(t)-\sqrt{\kappa_{\text{ex}}}\big[z_{\text{out,H}}^{a}(t)-z_{\text{out,H}}^{b}(t)\big],
\end{align}
\end{subequations}

\noindent where $\kappa = \kappa_\text{i} + \kappa_\text{ex}$ and,

\begin{subequations}
    \begin{align}
z_{\text{in,H}}^{a}(t)&=\frac{i}{\sqrt{2\pi}}\int_{-\infty}^{\infty}d\omega z_\text{H}^{a}(\omega)(t_{\text{in}})e^{-i\omega(t-t_{\text{in}})},\\z_{\text{in,H}}^{b}(t)&=\frac{i}{\sqrt{2\pi}}\int_{-\infty}^{\infty}d\omega z_\text{H}^{b}(\omega)(t_{\text{in}})e^{-i\omega(t-t_{\text{in}})},\\z_{\text{out,H}}^{a}(t)&=\frac{i}{\sqrt{2\pi}}\int_{-\infty}^{\infty}d\omega z_\text{H}^{a}(\omega)(t_{\text{out}})e^{-i\omega(t-t_{\text{out}})},\\z_{\text{out,H}}^{b}(t)&=\frac{i}{\sqrt{2\pi}}\int_{-\infty}^{\infty}d\omega z_\text{H}^{b}(\omega)(t_{\text{out}})e^{-i\omega(t-t_{\text{out}})}.
\end{align}
\end{subequations}

\noindent Using Eq. \ref{eqn:four_A_B_input_output_relation_source}, we obtain the IO relations for the bath $z^a$,
\begin{align}
    z_{\text{out,H}}^{a}(t)-z_{\text{in,H}}^{a}(t)=\sqrt{\kappa_{\text{ex}}}a_\text{H}(t)    
\end{align}
\noindent and similarly, we can compute for bath $z^b$,
\begin{align}
    z_{\text{out,H}}^{b}(t)-z_{\text{in,H}}^{b}(t)=\sqrt{\kappa_{\text{ex}}}b_\text{H}(t)
\end{align}
\noindent Assuming that the density matrix at $t=t_{\text{in}}$ is a product state between the molecular microtoroidal cavity and the continua,
\begin{align}
    \rho_{\text{total}}(t_{\text{in}})=\rho_{z^{a}}(t_{\text{in}})\otimes\rho(t_{\text{in}})\otimes\rho_{z^{b}}(t_{\text{in}}),
    \label{eqn:initial_state}
\end{align}

\noindent and that the driving occurs only from the $z^a$ bath,
\begin{subequations} \label{eqn:input_bath}
    \begin{align}
    \langle z_{\text{in,H}}^{a}(t)\rangle&\neq0 \label{eqn:input_bath_non_zero}
    \\
    \langle z_{\text{in,H}}^{b}(t)\rangle&=0 ,\label{eqn:input_bath_zero}
\end{align}
\end{subequations}
\noindent we get, after tracing over the continua,
\begin{align}
\frac{\partial X_\text{H}(t)}{\partial t}&=-i\big(\omega_{\text{ph}}+\beta\big)X_\text{H}(t)-\frac{\kappa}{2}X_\text{H}(t)+\sqrt{2}\sum_{i=1}^{N}g\left|g_{i}\right\rangle \left\langle e_{i}\right|(t)-\sqrt{\kappa_{\text{ex}}}\langle z_{\text{in,H}}^{a}(t)\rangle
\label{eqn:eom_A_H_t_final}
\end{align}
\noindent and 
\begin{align}
\frac{\partial Y_\text{H}(t)}{\partial t}&=-i\big(\omega_{\text{ph}}-\beta\big)Y_\text{H}(t)-\frac{\kappa}{2}Y_\text{H}(t)-\sqrt{\kappa_{\text{ex}}}\langle z_{\text{in,H}}^{a}(t)\rangle
\label{eqn:eom_B_H_t_final}
\end{align}
\noindent Eq. \ref{eqn:eom_A_H_t_final} can be rewritten as
\begin{align}
\frac{\partial X_\text{H}(t)}{\partial t}=-\frac{i}{\hbar}\big[X_\text{H}(t),\tilde{H}_\text{H}^{X}(t)\big]
\label{eqn:A_H_eom}
\end{align}
\noindent which allows us to conclude that the effective time-dependent Hamiltonian (in the Schrodinger picture) governing the molecular microtoroid cavity system is 
\begin{align}
    \tilde{H}^{X}(t)=H^{X'}+H_\text{int}^{X}(t).
\end{align}
\noindent In the absence of drive, the molecular microtoroidal cavity system obeys the effective non-Hermitian Hamiltonian,
\begin{align}
H^{X'}&=\hbar\Big[\omega_{\text{ph}}+\beta-i\frac{\kappa}{2}\Big]X^{\dagger}X+\sqrt{2}\sum_{i=1}^{N}\hbar gX^{\dagger}\left|g_{i}\right\rangle \left\langle e_{i}\right|+\sqrt{2}\sum_{i=1}^{N} \hbar gX\left|e_{i}\right\rangle \left\langle g_{i}\right|
\label{eqn:H_effective_X_mode}
\end{align}
\noindent while the time-dependent drive of the cavity due to light coupling from the bath $z^a$ is 
\begin{align}
    H_{\text{int}}^{X}(t)=-i\hbar\sqrt{\kappa_{\text{ex}}}\langle z_{\text{in,H}}^{a}(t)\rangle X^{\dagger}+h.c.
\end{align}
\noindent Similarly, we can use Eq. \ref{eqn:eom_B_H_t_final} and write the following set of equations that describes the effective Hamiltonian:
\begin{subequations}
    \begin{align}
\frac{\partial Y_\text{H}(t)}{\partial t}&=-\frac{i}{\hbar}\big[Y_\text{H}(t),\tilde{H}_\text{H}^{Y}(t)\big],
\\
\tilde{H}_\text{H}^{Y}(t)&=H_\text{H}^{Y'}+H_\text{int,H}^{Y}(t),
\\
H_\text{H}^{Y'}&=\hbar\Big[\omega_{\text{ph}}-\beta-i\frac{\kappa}{2}\Big]Y^{\dagger}Y,
\\
H_{\text{int,H}}^{Y}(t)&=-i\hbar\sqrt{\kappa_{\text{ex}}}\langle z_{\text{in,H}}^{a}(t)\rangle Y^{\dagger}+h.c.
\end{align}
\end{subequations}
\noindent Equipped with this formalism, we are interested in computing the following spectroscopic observables:
\begin{subequations} \label{eqn:spectroscopic variables}
\begin{align}
   \nonumber T(\omega)&=\frac{|\langle z_{\text{out,H}}^{a}(\omega)\rangle|^{2}}{|\langle z_{\text{in,H}}^{a}(\omega)\rangle|^{2}}=\frac{|\langle z_{\text{in,H}}^{a}(\omega)\rangle+\sqrt{\kappa_{\text{ex}}}\langle a_\text{H}(\omega)\rangle|^{2}}{|\langle z_{\text{in,H}}^{a}(\omega)\rangle|^{2}}\\
    &=\frac{|\langle z_{\text{in,H}}^{a}(\omega)\rangle+\frac{\sqrt{\kappa_{\text{ex}}}}{2}\Big[\langle X_\text{H}(\omega)\rangle+\langle Y_\text{H}(\omega)\rangle\Big]|^{2}}{|\langle z_{\text{in,H}}^{a}(\omega)\rangle|^{2}},\\
\nonumber R(\omega)&=\frac{|\langle z_{\text{out,H}}^{b}(\omega)\rangle|^{2}}{|\langle z_{\text{in,H}}^{a}(\omega)\rangle|^{2}}=\frac{|\sqrt{\kappa_{\text{ex}}}\langle b_\text{H}(\omega)\rangle|^{2}}{|\langle z_{\text{in,H}}^{a}(\omega)\rangle|^{2}}\\
&=\frac{|\frac{\sqrt{\kappa_{\text{ex}}}}{2}\Big[\langle X_\text{H}(\omega)\rangle-\langle Y_\text{H}(\omega)\rangle\Big]|^{2}}{|\langle z_{\text{in,H}}^{a}(\omega)\rangle|^{2}},
\\
A(\omega)&=1-T(\omega)-R(\omega)
\end{align}
\end{subequations}

\noindent where the traces above are carried out with respect to the initial state [Eq. \ref{eqn:initial_state}], $\langle\cdot\rangle=\text{Tr}\big[\cdot\rho_\text{total}(t_\text{in})\big]$, and in particular, $\langle \alpha_\text{H}(\omega)\rangle = \text{Tr} \big[ \alpha_\text{H}(\omega)\rho(t_\text{in})\big]$ with $\alpha=X,Y$ depends only on the initial state of the molecular microtoroidal cavity system. We have also used Eq. \ref{eqn:input_bath} and the Fourier transform convention, $f(\omega)=-i\int_{-\infty}^\infty dt e^{i\omega t}f(t)$. Equation \ref{eqn:spectroscopic variables} reveal that all the relevant spectroscopic observables can be obtained once $\langle X_\text{H} (\omega)\rangle$ and  $\langle Y_\text{H} (\omega)\rangle$ are known. 

\section{Kubo linear response}

\noindent Hereafter, we set $t_{\text{in}}=0$. As $H$ contains anharmonic terms, the evaluation of $\langle X(\omega)\rangle$ cannot be performed exactly. Hence, we carry out a Dyson expansion in $H_{\text{int}}$ for each of the Heisenberg operators in Eq. \ref{eqn:A_H_eom} and solve for $X^{(n)}_\text{H} (t)$ up to lowest nonvanishing order $n$.~\cite{yuen2023linear}

\noindent Starting at zeroth-order, $\mathcal{O}(H_\text{int}^0)$,
\begin{align}
    \frac{\partial X_{\text{H}}^{(0)}(t)}{\partial t}-\frac{i}{\hbar}\big[H^{X'},X_\text{H}^{(0)}(t)\big]=0
\end{align}
\noindent can be solved by
\begin{align}
    X_\text{H}^{(0)}(t)&=e^{iH^{X'}t/\hbar}Xe^{-iH^{X'}t/\hbar}.
\end{align}
\noindent Recall our assumption that the initial molecular microtoroidal cavity state is a product state between photon and molecular degrees of freedom, $\rho(0)=\rho_\text{ph}\otimes\rho_\text{mol}$, 
\begin{align}
    e^{-iH^{X'}t/\hbar}\rho(0)&\stackrel{t\to\infty}{=}\left|0\right\rangle \left\langle \varphi_{\text{ph}}\right|\otimes\rho_{\text{mol}},
\end{align}
\noindent where $\left|\varphi_{\text{ph}}\right\rangle $ is a photonic state. Then

\begin{subequations}
\begin{align}
    \langle X_{\text{H}}^{(0)}(t)\rangle&=\text{Tr}\big[e^{iH^{X'}t/\hbar}Xe^{-iH^{X'}t/\hbar}\rho(t_{\text{in}})\big]\\&\stackrel{t\to\infty}{=}0
\end{align}
\end{subequations}
\noindent which makes sense since any transient photonic amplitude will vanish due to photon escape.
Similarly, at $\mathcal{O}(H_{\text{int}})$, we have
\begin{align}
    \frac{\partial X_{\text{H}}^{(1)}(t)}{\partial t}-\frac{i}{\hbar}\big[H^{X'},X_{\text{H}}^{(1)}(t)\big]=-\sqrt{\kappa_{\text{ex}}}\langle z_{\text{in,H}}^{a}(t)\rangle.
\end{align}
\noindent This is a first-order inhomogeneous differential equation that can be solved with Green's function methods. We define
\begin{align}
    G^{XR}(t)&=\Theta(t)\Big[X_{\text{H}}^{(0)}(t),X_{\text{H}}^{(0)\dagger}(0)\Big]
\end{align}
\noindent which solves
\begin{align}
    \frac{\partial G^{XR}(t-t')}{\partial t}-\frac{i}{\hbar}\Big[H^{X'},D^{X}(t-t')\Big]=\delta(t-t')
\end{align}
\noindent where we need an additional assumption: the trace is performed over an initial state $\rho(t_\text{in})$ that is stationary with respect to $H^{X'}$; thus, it contains no photons. We readily obtain the Kubo linear response formula

\begin{subequations}
\begin{align}
    \langle X^{(1)}(t)\rangle&=-\sqrt{\kappa_{\text{ex}}}\int_{-\infty}^{\infty}dt_{1}\langle z_{\text{in}}^{a}(t-t_{1})\rangle D^{X}(t_{1})\\&=-\sqrt{\kappa_{\text{ex}}}\int_{-\infty}^{\infty}dt_{1}\langle z_{\text{in}}^{a}(t-t_{1})\rangle\Theta(t_{1})\langle\big[X_{\text{H}}^{(0)}(t_{1}),X^{\dagger}\big]\rangle,
\label{eqn:kubo_linear_response}
\end{align}
\end{subequations}
\noindent where $D^{X}(t)=\langle G^{XR}(t)\rangle$ is the retarded Green's function. Importantly, Eq. \ref{eqn:kubo_linear_response} has the form of a convolution,
\begin{align}
    \langle X(\omega)\rangle&=\langle X_{\text{H}}^{(1)}(\omega)\rangle=-i\sqrt{\kappa_{\text{ex}}}\langle z_{\text{in,H}}^{a}(\omega)\rangle D^{X}(\omega),
    \label{eqn:expression_a_omega}
\end{align}
\noindent where $D^X(\omega)$, according to the Fourier transform convention $f(\omega)=-i\int_{-\infty}^\infty dt e^{i\omega t}f(t)$, reads
\begin{align}
    D^{X}(\omega)&=-i\int_{-\infty}^{\infty}dte^{i\omega t}\Theta(t)\langle\big[e^{iH^{X'}t/\hbar}Xe^{-iH^{X'}t/\hbar},X^{\dagger}\big]\rangle.
\end{align}
\noindent Incidentally, given the non-Hermiitan nature of $H^{X'}$ due to photon leakage, $\rho(0)$ cannot contain photons, so one of the terms in the commutator is superfluous and the final propagator can be replaced, $H^{X'} \to H^{X'}$,
\begin{align}
    D^{X}(\omega)=-i\int_{-\infty}^{\infty}dte^{i\omega t}\Theta(t)\langle e^{iH^{X}t/\hbar}Xe^{-iH^{X'}t/\hbar}X^{\dagger}\rangle.
    \label{eqn:D^A_formula}
\end{align}
\noindent Similarly, for the anti-symmetric mode $Y$,
\begin{subequations}
\begin{align}
\langle Y(\omega)\rangle&=-i\sqrt{\kappa_{\text{ex}}}\langle z_{\text{in,H}}^{a}(\omega)\rangle D^{Y}(\omega),\label{eqn:exp_B_input}\\
    D^{Y}(\omega)&=-i\int_{-\infty}^{\infty}dte^{i\omega t}\Theta(t)\langle e^{iH^{Y}t/\hbar}Ye^{-iH^{Y'}t/\hbar}Y^{\dagger}\rangle.
    \label{eqn:D^B_formula}
\end{align}
\end{subequations}
\noindent Equation \ref{eqn:D^A_formula} and \ref{eqn:D^B_formula} can be fed into Eq. \ref{eqn:expression_a_omega} and \ref{eqn:exp_B_input}. Using Eqs. \ref{eqn:spectroscopic variables},

\begin{subequations}
\begin{align}
    T(\omega) &=1+\kappa_{\text{ex}}\Im\big[D^{X}(\omega)+D^{Y}(\omega)\big]+\frac{\kappa_{\text{ex}}^{2}}{4}\Big|D^{X}(\omega)+D^{Y}(\omega)\Big|^{2} \\ 
    R(\omega) &=\frac{\kappa_{\text{ex}}^{2}}{4}\Big|D^{X}(\omega)-D^{Y}(\omega)\Big|^{2} \\
    A(\omega) &= \sum_{\alpha = X,Y} A^{\alpha}(\omega), \\
    A^{\alpha}(\omega) &= -\kappa_{\text{ex}}\Im\big[D^{\alpha}(\omega)\big] - \frac{\kappa_{\text{ex}}^{2}}{2}\big|D^{\alpha}(\omega)\big|^{2}.
\label{eqn:spectroscopic_variables_final}
\end{align}
\end{subequations}
We summarise the results of this section in Table \ref{tab:spec}.

\begin{table}
	\centering
	% Captions go above tables
	\caption{\textbf{Linear response of a microtoroidal resonator coupled to $N$ molecules.} 
		This table presents the expressions for transmission, reflection, and absorption in a microtoroidal resonator interacting with $N$ molecules.}
	\label{tab:spec} % give each table a logical label name
	
	\begin{tabular}{lc} % two columns, left-aligned first column
		\\
		\hline
		\textbf{Response} & \textbf{Expression} \\
		\hline
		Transmission & $T(\omega) = 1+\kappa_{\text{ex}}\Im\big[D^{X}(\omega)+D^{Y}(\omega)\big]+\frac{\kappa_{\text{ex}}^{2}}{4}\Big|D^{X}(\omega)+D^{Y}(\omega)\Big|^{2}$ \\  
		Reflection & $R(\omega) = \frac{\kappa_{\text{ex}}^{2}}{4}\Big|D^{X}(\omega)-D^{Y}(\omega)\Big|^{2}$ \\  
		Absorption & $A(\omega) = -\kappa_{\text{ex}}\Im\big[D^{X}(\omega)\big] - \frac{\kappa_{\text{ex}}^{2}}{2} \big|D^{X}(\omega)\big|^{2} - \kappa_{\text{ex}}\Im\big[D^{Y}(\omega)\big] - \frac{\kappa_{\text{ex}}^{2}}{2} \big|D^{Y}(\omega)\big|^{2}$ \\  
		\hline
	\end{tabular}
\end{table}

% \begin{table}[h]
%     \centering
%     \renewcommand{\arraystretch}{2
%     } % Increase row height
%     \begin{tabular}{|c|c|} \hline 
        
%         \textbf{Linear response}& \textbf{Microtoroid resonator coupled to $N$ molecules}\\ \hline 
        
%         Transmission & $T(\omega) = 1+\kappa_{\text{ex}}\Im\big[D^{X}(\omega)+D^{Y}(\omega)\big]+\frac{\kappa_{\text{ex}}^{2}}{4}\Big|D^{X}(\omega)+D^{Y}(\omega)\Big|^{2}$\\ \hline  
        
%         Reflection & $R(\omega) = \frac{\kappa_{\text{ex}}^{2}}{4}\Big|D^{X}(\omega)-D^{Y}(\omega)\Big|^{2}$\\ \hline 
        
%         Absorption & $A(\omega) = -\kappa_{\text{ex}}\Im\big[D^{X}(\omega)\big] - \frac{\kappa_{\text{ex}}^{2}}{2} \big|D^{X}(\omega)\big|^{2} - \kappa_{\text{ex}}\Im\big[D^{Y}(\omega)\big] - \frac{\kappa_{\text{ex}}^{2}}{2} \big|D^{Y}(\omega)\big|^{2}$\\ \hline
        
%     \end{tabular}
%     \caption{Linear response of a microtoroidal resonator coupled to $N$ molecules}
%     \label{tab:spec}
% \end{table}

\section{Absorption spectra for arbitrary $N$ molecules coupled to microtoroid resonator}

\noindent According to Eq. \ref{eqn:spectroscopic_variables_final}, there are two sources of absorption, 
\begin{subequations}
\begin{align}
 A^{X}(\omega)&=-\kappa_{\text{ex}}\Im\big[D^{X}(\omega)\big]-\frac{\kappa_{\text{ex}}^{2}}{2}\big|D^{X}(\omega)\big|^{2}\\A^{Y}(\omega)&=-\kappa_{\text{ex}}\Im\big[D^{Y}(\omega)\big]-\frac{\kappa_{\text{ex}}^{2}}{2}\big|D^{Y}(\omega)\big|^{2}
 \label{eqn:A^Y_final}
\end{align}
\end{subequations}

\subsection{Absorption via anti-symmetric cavity mode}

\noindent The absorption due to molecules is present only in $A^{X}(\omega)$ as only $X$ couples with the molecules. On the other hand, $A^{Y}(\omega)$ accounts for the absorption due to the intrinsic loss in the microtoroidal cavity. Thus, it is straightforward to compute $D^{Y}(\omega)$ using Eq. \ref{eqn:D^B_formula}

\begin{subequations} 
\begin{align}
 D^{Y}(\omega)&=-i\int_{-\infty}^{\infty}dte^{i\omega t}\Theta(t)\langle e^{\frac{i}{\hbar}H^{Y'}t}Ye^{-\frac{i}{\hbar}H^{Y'}t}Y^{\dagger}\rangle\\&=\frac{1}{\omega-(\omega_{\text{ph}}-\beta)+i\big(\frac{\kappa_{i}+\kappa_{\text{ex}}}{2}\big)}.
 \label{eqn:D_Y_final}
\end{align}
\end{subequations}

\noindent Feeding Eq.~\ref{eqn:D_Y_final} into Eq.~\ref{eqn:A^Y_final} leads to,

\begin{align}
 A^{Y}(\omega)&=\frac{\kappa_{\text{ex}}\kappa_{i}}{\big[\omega-(\omega_{\text{ph}}-\beta)\big]^{2}+\big(\frac{\kappa_{i}+\kappa_{\text{ex}}}{2}\big)^{2}}
\end{align}

\subsection{Absorption via symmetric cavity mode}

\noindent The absorption due to the symmetric mode $X$ can be computed using 

\begin{align}
D^{X}(\omega)=-i\int_{-\infty}^{\infty}dte^{i\omega t}\Theta(t)\langle e^{\frac{i}{\hbar}H^{X'}t}Xe^{-\frac{i}{\hbar}H^{X'}t}X^{\dagger}\rangle,
\end{align}
where $\langle\dots\rangle$ is taken with respect to the zero temperature photon-less
initial state. Performing the Fourier transform, we obtain~\cite{MukamelBook}, 
\begin{align}
D^{X}(\omega)=\langle XG^{X}(\omega)X^{\dagger}\rangle,
\label{eqn:G^X(w)}
\end{align}
where $G^{X}(\omega)=\frac{1}{\omega-\tilde{H}^{X}+i0^{+}}$ is the total Green's function corresponding to the full Hamiltonian of the symmetric mode coupled to the $N$ molecules.

\section{Microresonator coupled to single molecule}

\noindent The Green's function of the effective Hamiltonian for the symmetric optical mode of the microresonator in Eq.~\ref{eqn:H_effective_X_mode} is $G^X(\omega)=1/(\omega-\tilde{H}^X+i0^+)$. Now, the Dyson equation,
\begin{align}
G^{X}(\omega)=g(\omega)+g(\omega)VG^{X}(\omega).
\end{align}
where, for single molecule, $g(\omega)=1/(\omega-(\omega_{ph}+\beta)+i\frac{\kappa}{2})$ and $V=\tilde{g}X^\dagger|g\rangle\langle e|+h.c.$. Here, we considered $\tilde{g}=\sqrt{2}g$. Now, the Dyson series is
\begin{align}
G^{X}(\omega)=g(\omega)+g(\omega)Vg(\omega)+g(\omega)Vg(\omega)Vg(\omega)+\dots
\end{align}
As shown in Eq.~\ref{eqn:G^X(w)}, $\langle XG^{X}(\omega)X^{\dagger}\rangle$ is the total projection
of the total Green's function onto the symmetric mode of the microtoroid.
We will now compute the Dyson series term by term. Note that for this
example, we have the microresonator off-resonant to the electronic transition
of the molecule, $\Delta=|(\omega_{\text{ph}}+\beta)-\omega_{e}|$. Since the light-matter
coupling $g\ll\Delta$, we are allowed to truncate the Dyson series~\cite{tannor2007introduction}.

\begin{figure*}[ht!]
     
        \includegraphics[width=0.9\textwidth]{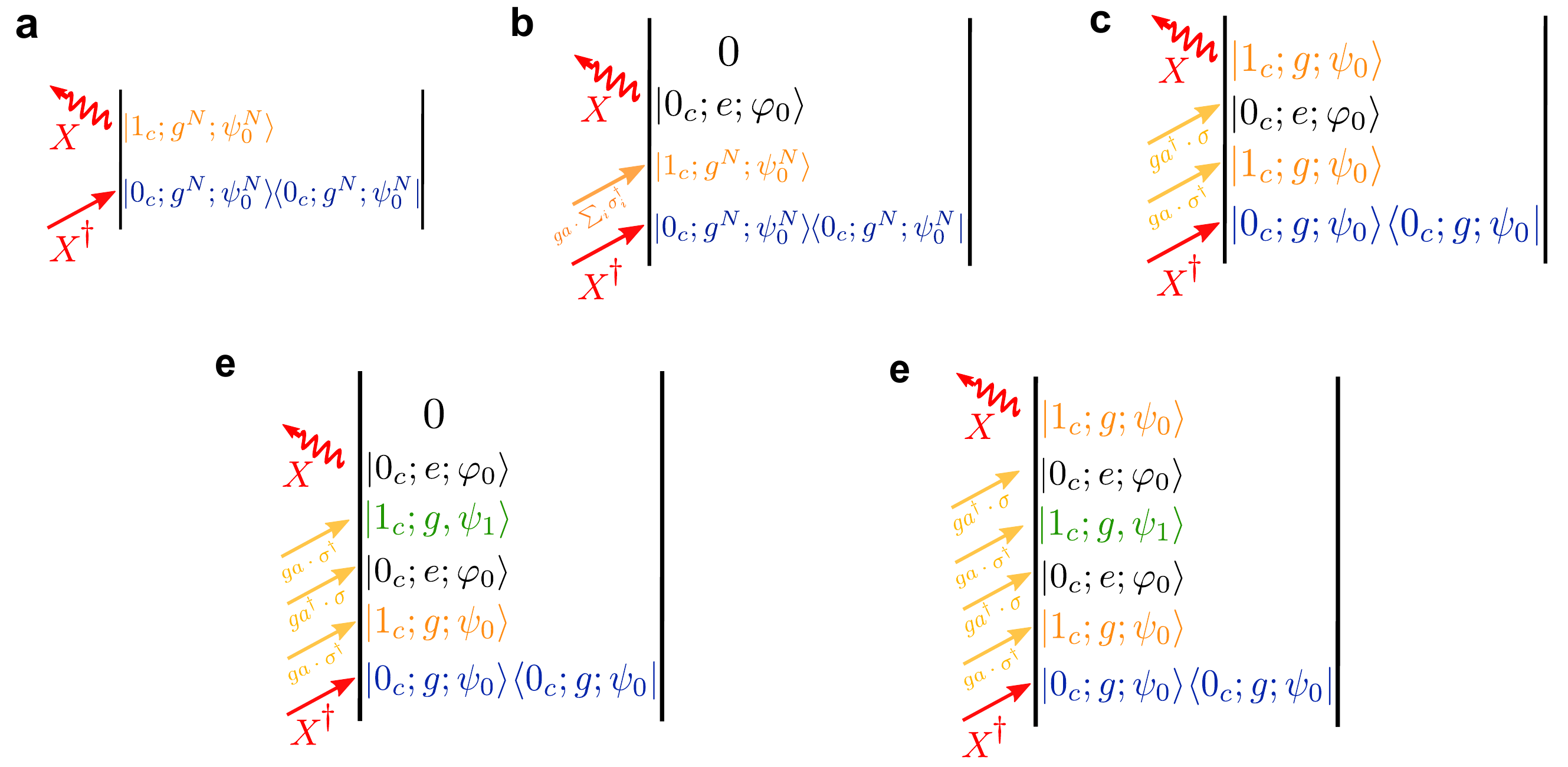}
        
\caption{\textbf{Double-Sided Feynman diagrams for the single molecule coupled to the high-$Q$ microresonator up to fourth order}: \textbf{a} Zeroth order, \textbf{b} first order, \textbf{c} second order, \textbf{d} third order, \textbf{e} fourth order terms in the light-matter couplings. Notice that the odd-order terms in the Dyson expansion vanish.}
\label{fig:single_molecule_FDSD}
\end{figure*}

\medskip{}
\noindent We have the zeroth order term $\langle XG^{X}(\omega)X^{\dagger}\rangle=\langle Xg(\omega)X^{\dagger}\rangle=\frac{1}{\omega-(\omega_{\text{ph}}+\beta)+i\kappa/2}$
which is the Green's function of the bare cavity. For the higher-order
terms of the Dyson expansion, we can use the technique of double-sided
Feynman diagrams (DSFDs)~\cite{MukamelBook}.
Note that the restriction to the linear response regime, conservation
of excitation number by the Hamiltonian, and a photon-less zero temperature initial
state only allows access to one of the sides of the diagram (we choose
it to be the ket side by convention). Fig.~\ref{fig:single_molecule_FDSD} shows
the DSFDs up to the fourth order term in $V$ for a representative
molecule with three levels: the ground electronic state with two vibrational
levels $|g,\psi_{0}\rangle$ and $|g,\psi_{1}\rangle$, and the excited
electronic state with one vibrational level $|e,\varphi_{0}\rangle$.
We notice that the first and the third order terms in the series are
zero. This is a known observation for subspace-projected Green's
functions for coupled systems~\cite{dittes2000decay, MukamelBook}. 
We obtain the expressions for the second and fourth-order terms
(where $D_{(n)}^{X}(\omega)$ represents the $n^{\text{th }}$order term
in the Dyson series) as
\begin{align}
   \nonumber D_{(2)}^{X}(\omega)&=\frac{1}{\omega-(\omega_{\text{ph}}+\beta)+i\kappa/2}g\langle\psi_{0}|\varphi_{0}\rangle\frac{1}{\omega-\omega_{e}+i\gamma/2}g\langle\varphi_{0}|\psi_{0}\rangle\frac{1}{\omega-(\omega_{\text{ph}}+\beta)+i\kappa/2}\\ 
    &=g^{2}\Big[D_{(0)}^{X}(\omega)\Big]^{2}\chi^{(1)}(\omega),
\end{align}

and,
\small
\begin{align}
    \nonumber D_{(4)}^{X}(\omega)&=\frac{1}{\omega-(\omega_{\text{ph}}+\beta)+i\kappa/2}g\langle\psi_{0}|\varphi_{0}\rangle\frac{1}{\omega-\omega_{e}+i\gamma/2}g\langle\varphi_{0}|\psi_{0}\rangle\frac{1}{\omega-(\omega_{\text{ph}}+\beta)+i\kappa/2}\\
    \nonumber &\quad\times g\langle\psi_{0}|\varphi_{0}\rangle\frac{1}{\omega-\omega_{e}+i\gamma/2}g\langle\varphi_{0}|\psi_{0}\rangle\frac{1}{\omega-(\omega_{\text{ph}}+\beta)+i\kappa/2}\\
    \nonumber &\quad+\frac{1}{\omega-(\omega_{\text{ph}}+\beta)+i\kappa/2}g\langle\psi_{0}|\varphi_{0}\rangle\frac{1}{\omega-\omega_{e}+i\gamma/2}g\langle\varphi_{0}|\psi_{1}\rangle\frac{1}{\omega-((\omega_{\text{ph}}+\beta)+\omega_{g})+i(\kappa+\gamma_{vib})/2}\\
    \nonumber &\quad\times g\langle\psi_{1}|\varphi_{0}\rangle\frac{1}{\omega-\omega_{e}+i\gamma/2}g\langle\varphi_{0}|\psi_{0}\rangle\frac{1}{\omega-(\omega_{\text{ph}}+\beta)+i\kappa/2}.\\
     &=g^{4}\Big[D_{(0)}^{X}(\omega)\Big]^{3}\Big[\chi^{(1)}(\omega)\Big]^{2}+R_{\text{vib}}(\omega)
\end{align}
\normalsize

\noindent where, in general, $\chi^{(1)}(\omega)=-\lim_{\gamma\to0^{+}}\sum_{n=0}\frac{|\left\langle \psi_{0}\big|\varphi_{n}\right\rangle |^{2}}{\omega-\omega_{e,n}+i\frac{\gamma}{2}}$
is the linear susceptibility of the molecule representing the cavity
mediated Rayleigh-type scattering processes through which energy is
dissipated into the molecular bath. And, the generalized form of $R_{\text{vib}}(\omega)$
is 
\begin{align}
R_{\text{vib}}(\omega)=g^{4}\left[D_{(0)}^{X}(\omega)\right]^{2}\sum_{m=1}^{M_{g}}\frac{\left\langle \psi_{0}\right|G_{\text{ex}}(\omega)\left|\psi_{m}\right\rangle \left\langle \psi_{m}\right|G_{\text{ex}}(\omega)\left|\psi_{0}\right\rangle }{\omega-\bigl((\omega_{\text{ph}}+\beta)+\omega_{g,m}\bigr)+i\,\tfrac{(\kappa+\gamma_{\text{vib}})}{2}}
\end{align}
 Here, $\ensuremath{G_{\text{ex}}(\omega)=\sum_{m}\frac{|\varphi_{m}\rangle\langle\varphi_{m}|}{(\omega-\omega_{e,m}+i\frac{\gamma}{2})}}$
is the excited state Green's function of the molecule. Now, $R_{\text{vib}}(\omega)$
under standard approximations can be expressed in terms of Stokes and
conditional anti-Stokes cross-section.

\subsection{Stokes cross-section}

\noindent Considering Real Franck-Condon Factors and $\gamma\to0$,
\begin{align}
R_{\text{vib}}(\omega)\approx g^{4}\Big[D_{(0)}^{X}(\omega)\Big]^{2}\sum_{m=1}^{M_{g}}\frac{\Big|\left\langle \psi_{0}\right|G_{e}(\omega)\left|\psi_{m}\right\rangle \Big|^{2}}{\omega-((\omega_{\text{ph}}+\beta)+\omega_{g,m})+i\frac{(\kappa+\gamma_{\text{vib}})}{2}}
\end{align}
Now, $R_{\text{vib}}(\omega)$ can be expressed in terms of Stokes
Raman cross-section,
\begin{align}
R_{\text{vib}}(\omega)\approx g^{4}\Big[D_{(0)}^{X}(\omega)\Big]^{2}S_{\text{Raman}}(\omega_{L}=\omega,\omega_{S}=(\omega_{\text{ph}}+\beta))
\end{align}
where, $S_{\text{Raman}}(\omega_{L}=\omega,\omega_{S}=(\omega_{\text{ph}}+\beta))$
is the traditional Stokes Raman cross-section~\cite{MukamelBook, tannor2007introduction, schatz2002quantum}.

\subsection{Conditional anti-Stokes cross-section}

\noindent Considering Real Franck-Condon Factors and $\gamma\to0$, $R_{\text{vib}}(\omega)$
around $(\omega_{\text{ph}}+\beta)$ is
\begin{align}
R_{\text{vib}}(\omega)\approx g^{4}\Big[D_{(0)}^{X}(\omega)\Big]^{2}\sum_{m=1}^{M_{g}}\frac{\Big|\left\langle \psi_{0}\right|G_{e}((\omega_{\text{ph}}+\beta))\left|\psi_{m}\right\rangle \Big|^{2}}{\omega-((\omega_{\text{ph}}+\beta)+\omega_{g,m})+i\frac{(\kappa+\gamma_{\text{vib}})}{2}}
\end{align}
Now, $R_{\text{vib}}(\omega)$ can be expressed in terms of Stokes
Raman cross-section,
\begin{align}
R_{\text{vib}}(\omega)\approx g^{4}\Big[D_{(0)}^{X}(\omega)\Big]^{2}S_{\text{Raman}}^{C}(\omega_{L}=(\omega_{\text{ph}}+\beta),\omega_{AS}=\omega)
\end{align}
 where, $S_{\text{Raman}}(\omega_{L}=\omega,\omega_{S}=(\omega_{\text{ph}}+\beta))$
is the conditional anti-Stokes Raman cross-section where the final
state of the molecule is constrained to be in the global ground state~\cite{MukamelBook, tannor2007introduction, schatz2002quantum}.

\subsection{Underlying Stokes and anti-Stokes mechanism}

\noindent Thus, $R_{\text{vib}}(\omega)$ near the cavity peak can be expressed
in terms of the underlying mechanism involving Stokes and subsequent
anti-Stokes process
\begin{align}
R_{\text{vib}}(\omega)\approx g^{4}\Big[D_{(0)}^{X}(\omega)\Big]^{2}\sqrt{S_{\text{Raman}}(\omega_{L}=\omega,\omega_{S}=(\omega_{\text{ph}}+\beta))}\sqrt{S_{\text{Raman}}^{C}(\omega_{L}=(\omega_{\text{ph}}+\beta),\omega_{AS}=\omega)}
\end{align}

\section{Exact solution for photon Green's function of a single molecule coupled to microresonator} \label{sec:exact_soln_single_molecule}

\noindent It turns out that for the single molecule case, we can obtain an exact expression for the photon Green's function. For simplicity, we assume all the Franck-Condon factors to be $1$ and write down all the higher-order terms in the Dyson expansion,

\begin{align}
\frac{1}{\omega-(\omega_{\text{ph}}+\beta)+i\kappa/2}=A,\quad\frac{1}{\omega-\omega_{e}+i\gamma/2}=B^ {},\quad\frac{1}{\omega-((\omega_{\text{ph}}+\beta)+\omega_{g})+i(\kappa+\gamma_{vib})/2}=C^ {}.
\end{align}
Using a straightforward combinatorial counting of all the nonzero
Feynman paths contributing to $\langle XG^{X}(\omega)X^{\dagger}\rangle$
at a given order, we obtain

\begin{subequations} \label{eqn:all_D_single_molecule}
\begin{align}
D_{(0)}^{X}(\omega) & =A,\\
D_{(0)}^{X}(\omega) & =\big[AgBg\big]A,\\
D_{(4)}^{X}(\omega) & =\big[AgBg\big]^{2}A+Ag(BgCg)BgA,\\
D_{(6)}^{X}(\omega) & =\big[AgBg\big]^{3}A+Ag(BgCg)^{2}BgA+AgBgCgBgAgBgA+AgBgAgBgCgBg,
\end{align}
\end{subequations}

\noindent and for arbitrary order, 

\begin{align}
D_{(2k)}^{X}(\omega) & =\sum_{n=0}^{\infty}\sum_{m_{n}=0}^{\infty}\cdots\sum_{m_{1}=0}^{\infty}\bigg(Ag\bigg[(BgCg)^{m_{1}}\bigg]Bg\bigg)\bigg(Ag\bigg[(BgCg)^{m_{2}}\bigg]Bg\bigg)\dots\bigg(Ag\bigg[(BgCg)^{m_{n}}\bigg]Bg\bigg)A
\end{align}

\noindent such that $\sum_{i=1}^{n}m_{i}+n=k$ with $n\neq0$. 
 Summing over
all the terms of this convergent series, we obtain

\begin{subequations}
  \begin{align}
\nonumber \langle XG^{X}(\omega)X^{\dagger}\rangle & =\sum_{n=0}^{\infty}\sum_{m_{n}=0}^{\infty}\cdots\sum_{m_{1}=0}^{\infty}\bigg(Ag\bigg[(BgCg)^{m_{1}}\bigg]Bg\bigg)\bigg(Ag\bigg[(BgCg)^{m_{2}}\bigg]Bg\bigg)\dots\bigg(Ag\bigg[(BgCg)^{m_{n}}\bigg]Bg\bigg)A\\ \nonumber
 & =\sum_{n=0}^{\infty}\bigg(Ag\bigg[\sum_{m=0}(BgCg)^{m}\bigg]Bg\bigg)^{n}A\\ 
 & =\sum_{n=0}^{\infty}\bigg(Ag\frac{1}{1-BgCg}Bg\bigg)^{n}A=\bigg(\frac{1}{1-Ag\frac{1}{1-BgCg}Bg}\bigg)A.
\end{align}  
\end{subequations}

\section{Dyson expansion for $N$ molecules coupled to the microresonator}

\begin{figure*}[ht!]
     
        \includegraphics[width=1\textwidth]{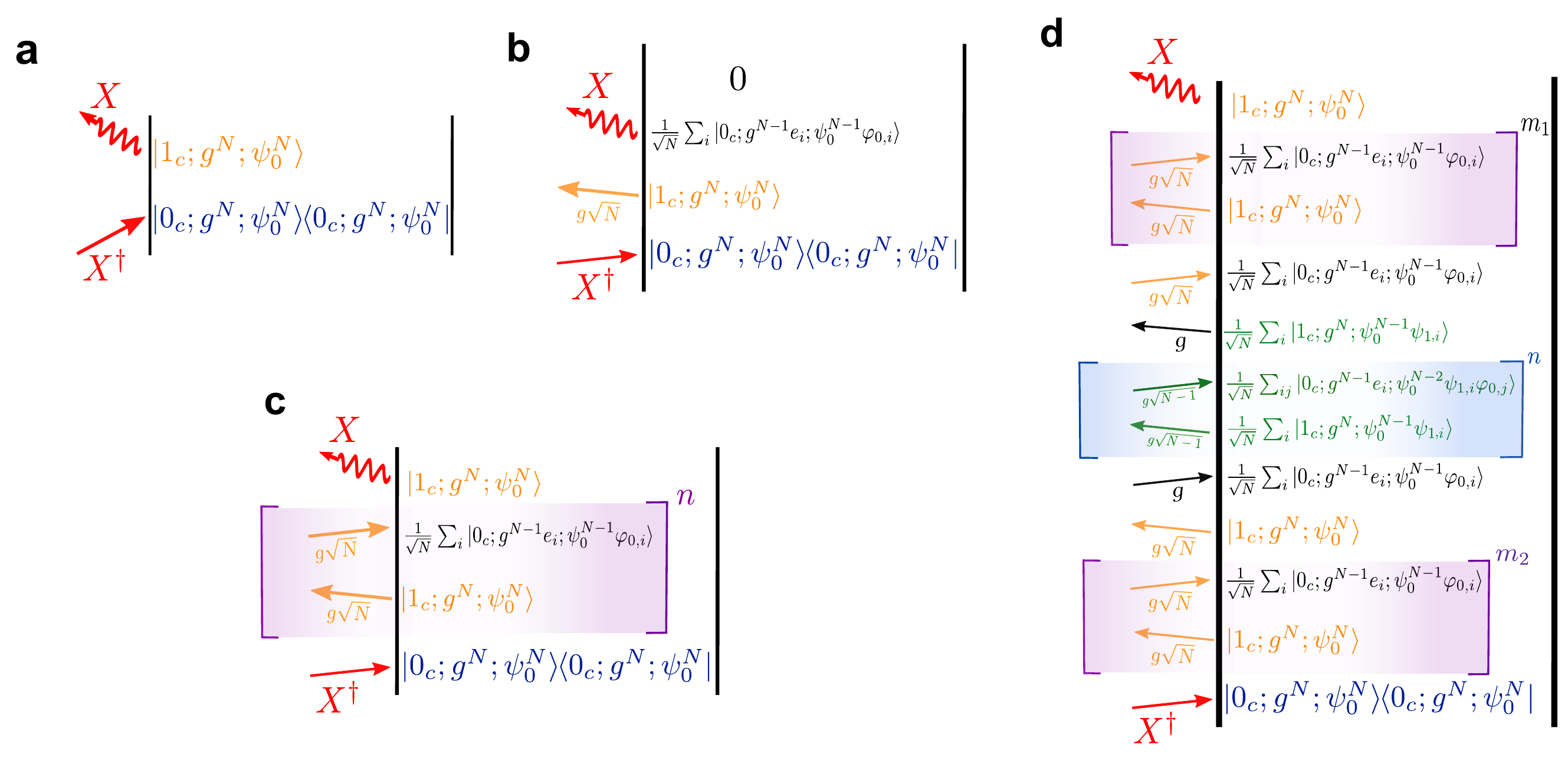}
        
\caption{\textbf{Double-Sided Feynman diagrams for the $N$ molecule coupled to the high-$Q$ microresonator up to second order in $g$}: \textbf{a} Zeroth order, \textbf{b} first order, (c) second order, (d) third order, (e) fourth order terms in the light-matter couplings. Notice that the odd-order terms in the Dyson expansion vanish. 
}
\label{fig:N_molecule_DSFD}
\end{figure*}

Let the zero temperature initial state be 
\begin{align}
|\Psi_{0}\rangle=|0_{c};g^{N};\psi_{0}^{N}\rangle,
\end{align}
where $|g^{N}\rangle=|ggg\dots g\rangle$ and $|\psi_{0}^{N}\rangle=|\psi_{0}\psi_{0}\dots\psi_{0}\rangle$
represents all the $N$ molecules in $|g,\psi_{0}\rangle$ state. Similar to the previous case, we can compute the terms of the Dyson series. Further, we also have the odd order terms in the Dyson expansion to be zero. Now, we write the terms of the series order-by-order using the DSFDs in Fig.~\ref{fig:N_molecule_DSFD}. We have, 

\begin{align}
D_{(2)}^{X(N)}(\omega)&=\frac{1}{\omega-(\omega_{\text{ph}}+\beta)+i\kappa/2}g\sqrt{N}\langle\psi_{0}|\varphi_{0}\rangle\frac{1}{\omega-\omega_{e}+i\gamma/2}g\sqrt{N}\langle\varphi_{0}|\psi_{0}\rangle\frac{1}{\omega-(\omega_{\text{ph}}+\beta)+i\kappa/2},\\\nonumber D_{(4)}^{X(N)}(\omega)&=\frac{1}{\omega-(\omega_{\text{ph}}+\beta)+i\kappa/2}\bigg(g\sqrt{N}\langle\psi_{0}|\varphi_{0}\rangle\frac{1}{\omega-\omega_{e}+i\gamma/2}g\sqrt{N}\langle\varphi_{0}|\psi_{0}\rangle\frac{1}{\omega-(\omega_{\text{ph}}+\beta)+i\kappa/2}\bigg)^{2}\\\nonumber&\quad+\frac{1}{\omega-(\omega_{\text{ph}}+\beta)+i\kappa/2}g\sqrt{N}\langle\psi_{0}|\varphi_{0}\rangle\frac{1}{\omega-\omega_{e}+i\gamma/2}g\langle\varphi_{0}|\psi_{1}\rangle  \\
&\quad\times \nonumber \frac{1}{\omega-((\omega_{\text{ph}}+\beta)+\omega_{g})+i(\kappa+\gamma_{vib})/2}g\langle\psi_{1}|\varphi_{0}\rangle\frac{1}{\omega-\omega_{e}+i\gamma/2}\\& \quad \times g\sqrt{N}\langle\varphi_{0}|\psi_{0}\rangle\frac{1}{\omega-(\omega_{\text{ph}}+\beta)+i\kappa/2},
 \\\nonumber D_{(6)}^{X(N)}(\omega)&=\frac{1}{\omega-(\omega_{\text{ph}}+\beta)+i\kappa/2}\bigg(g\sqrt{N}\langle\psi_{0}|\varphi_{0}\rangle\frac{1}{\omega-\omega_{e}+i\gamma/2}g\sqrt{N}\langle\varphi_{0}|\psi_{0}\rangle\frac{1}{\omega-(\omega_{\text{ph}}+\beta)+i\kappa/2}\bigg)^{3}+\\\nonumber&\quad+\frac{1}{\omega-(\omega_{\text{ph}}+\beta)+i\kappa/2}g\sqrt{N}\langle\psi_{0}|\varphi_{0}\rangle\bigg(\frac{1}{\omega-\omega_{e}+i\gamma/2}g\langle\varphi_{0}|\psi_{1}\rangle \\  \nonumber  & \quad 
 \frac{1}{\omega-((\omega_{\text{ph}}+\beta)+\omega_{g})+i(\kappa+\gamma_{vib})/2}g\langle\psi_{1}|\varphi_{0}\rangle\bigg)^{2}
 \frac{1}{\omega-\omega_{e}+i\gamma/2}
 \\\nonumber&\quad\times g\sqrt{N}\langle\varphi_{0}|\psi_{0}\rangle\frac{1}{\omega-(\omega_{\text{ph}}+\beta)+i\kappa/2}\\\nonumber&\quad+\frac{1}{\omega-(\omega_{\text{ph}}+\beta)+i\kappa/2}g\sqrt{N}\langle\psi_{0}|\varphi_{0}\rangle\frac{1}{\omega-\omega_{e}+i\gamma/2}g\langle\varphi_{0}|\psi_{1}\rangle
\\ \nonumber &\quad \times   
 \frac{1}{\omega-((\omega_{\text{ph}}+\beta)+\omega_{g})+i(\kappa+\gamma_{vib})/2}\\\nonumber&\quad\times\bigg(g\sqrt{N-1}\langle\psi_{0}|\varphi_{0}\rangle\frac{1}{\omega-(\omega_{e}+\omega_{g})+i(\gamma+\gamma_{vib})/2}\\\nonumber&\quad\times g\sqrt{N-1}\langle\varphi_{0}|\psi_{0}\rangle\frac{1}{\omega-((\omega_{\text{ph}}+\beta)+\omega_{g})+i(\kappa+\gamma_{vib})/2}\bigg)\\&\quad\times g\langle\psi_{1}|\varphi_{0}\rangle\frac{1}{\omega-\omega_{e}+i\gamma/2}g\sqrt{N}\langle\varphi_{0}|\psi_{0}\rangle\frac{1}{\omega-(\omega_{\text{ph}}+\beta)+i\kappa/2}
\end{align}
Now assuming $N$ is large, such that $g\sqrt{N-1}\approx g\sqrt{N}$,
we see that there are two emergent timescales in the problem: the collective timescale, $g\sqrt{N},$ that involves only the linear susceptibility of the molecular ensemble, and the single molecule timescale, $g$, that mediates the Raman processes. Since $g\sqrt{N}\gg g$, and from our experience with the single molecule case, we expect the Raman terms to be proportional to $g^4$ (which is essentially
$g^{2}N\times g^{2}$). Thus, the term in the Dyson expansion that accounts for $2k$ orders in collective timescale and $2$ orders in single molecule timescale (this can also be checked in a similar way using a combinatorial counting of Feynman paths Fig.~\ref{fig:N_molecule_DSFD},

\small
\begin{align}
\nonumber D_{(2k)}^{X(N)}(\omega)&=\frac{1}{\omega-(\omega_{\text{ph}}+\beta)+i\kappa/2}\bigg(g\sqrt{N}\langle\psi_{0}|\varphi_{0}\rangle\frac{1}{\omega-\omega_{e}+i\gamma/2}g\sqrt{N}\langle\varphi_{0}|\psi_{0}\rangle\frac{1}{\omega-(\omega_{\text{ph}}+\beta)+i\kappa/2}\bigg)^{k}\\\nonumber &+\sum_{n=0}^{k}\sum_{m_{2}=0}^{k}\sum_{m_{1}=0}^{k}\frac{1}{\omega-(\omega_{\text{ph}}+\beta)+i\kappa/2}\bigg(g\sqrt{N}\langle\psi_{0}|\varphi_{0}\rangle\frac{1}{\omega-\omega_{e}+i\gamma/2}g\sqrt{N}\langle\varphi_{0}|\psi_{0}\rangle\frac{1}{\omega-(\omega_{\text{ph}}+\beta)+i\kappa/2}\bigg)^{m_{1}}\\
\nonumber &\times\bigg[g\sqrt{N}\langle\psi_{0}|\varphi_{0}\rangle\frac{1}{\omega-\omega_{e}+i\gamma/2}g\langle\varphi_{0}|\psi_{1}\rangle\frac{1}{\omega-((\omega_{\text{ph}}+\beta)+\omega_{g})+i(\kappa+\gamma_{vib})/2}\\
\nonumber &\times\bigg(g\sqrt{N}\langle\psi_{0}|\varphi_{0}\rangle\frac{1}{\omega-(\omega_{e}+\omega_{g})+i(\gamma+\gamma_{vib})/2}g\sqrt{N}\langle\varphi_{0}|\psi_{0}\rangle\frac{1}{\omega-((\omega_{\text{ph}}+\beta)+\omega_{g})+i(\kappa+\gamma_{vib})/2}\bigg)^{n}\\
\nonumber &\times g\langle\psi_{1}|\varphi_{0}\rangle\frac{1}{\omega-\omega_{e}+i\gamma/2}g\sqrt{N}\langle\varphi_{0}|\psi_{0}\rangle\bigg]\frac{1}{\omega-(\omega_{\text{ph}}+\beta)+i\kappa/2}\\&\times\bigg(g\sqrt{N}\langle\psi_{0}|\varphi_{0}\rangle\frac{1}{\omega-\omega_{e}+i\gamma/2}g\sqrt{N}\langle\varphi_{0}|\psi_{0}\rangle\frac{1}{\omega-(\omega_{\text{ph}}+\beta)+i\kappa/2}\bigg)^{m_{2}}+\mathcal{O}(g^{4}),
\end{align}
\normalsize
\noindent such that $m_{1}+m_{2}+n+2=k$. Here, the reader can verify that the higher order terms in $g$, \emph{e.g.} $\mathcal{O}(g^6)$ for the case of $N$ molecules coupled to the resonator contain new vacuum-mediated features that are not present in the single-molecule case. Since $g$ is small, these features are even smaller than the sought-out Raman peaks. Thus, observing these peaks requires ultrahigh $Q$ cavities, which are a topic for future work. \\

\noindent Further, the key difference between the case of $N$ molecules and a single molecule coupled to the resonator is the emergence of a collective fast timescale that forbids truncation of terms involving powers of $g\sqrt{N}$ (as $g\sqrt{N}$ leads to a divergent series). Thus, this necessitates the exact resummation of these collective terms. The issue of such apparent divergences is well known in the literature of Quantum electrodyanmics (QED) and Quantum chromodyanmics (QCD) where several asymptotic techniques have been developed to do exact resummation. We use the method of Borel summation shown in the Appendix to resum the series. We obtain,

\small
\begin{align}
    \nonumber D_{(0)}^{X(N)}(\omega) =
    S_{1} & =\sum_{k=0}^{\infty}\frac{1}{\omega-(\omega_{\text{ph}}+\beta)+i\kappa/2}\bigg(g\sqrt{N}\langle\psi_{0}|\varphi_{0}\rangle\frac{1}{\omega-\omega_{e}+i\gamma/2}g\sqrt{N}\langle\varphi_{0}|\psi_{0}\rangle\frac{1}{\omega-(\omega_{\text{ph}}+\beta)+i\kappa/2}\bigg)^{k},\\
 & =\frac{1}{\omega-(\omega_{\text{ph}}+\beta)+i\kappa/2-g^{2}N\frac{|\langle\psi_{0}|\varphi_{0}\rangle|^{2}}{\omega-\omega_{e}+i\gamma/2}}
\end{align}
\normalsize
\noindent Similarly, using Borel summation,

\begin{align}
\nonumber D_{(2)}^{X(N)}(\omega)=S_{2}&=D_{(0)}^{X(N)}(\omega)\bigg[g\sqrt{N}\langle\psi_{0}|\varphi_{0}\rangle\frac{1}{\omega-\omega_{e}+i\gamma/2}g\langle\varphi_{0}|\psi_{1}\rangle\\&\quad\times\frac{1}{\omega-((\omega_{\text{ph}}+\beta)+\omega_{g})+i(\kappa+\gamma_{vib})/2}g\sqrt{N}\langle\psi_{0}|\varphi_{0}\rangle\\&\quad\times\frac{1}{\omega-((\omega_{\text{ph}}+\beta)+\omega_{vib})+i(\kappa+\gamma_{vib})/2-g^{2}N\frac{|\langle\psi_{0}|\varphi_{0}\rangle|^{2}}{\omega-(\omega_{e}+\omega_{vib})+i(\gamma+\gamma_{vib})/2}}g\langle\psi_{1}|\varphi_{0}\rangle\\&\quad\times\frac{1}{\omega-\omega_{e}+i\gamma/2}g\sqrt{N}\langle\varphi_{0}|\psi_{0}\rangle\bigg]D_{(0)}^{X(N)}(\omega).
\end{align}

\noindent This is the final expression we obtain for the photon Green's function
up to $g^{2}$. We have, 

\begin{align}
D^{X(N)}(\omega)=D_{(0)}^{X(N)}(\omega)+D_{(2)}^{X(N)}(\omega)+\mathcal{O}(g^{4}).
\end{align}
Here, 
\small
\begin{align}
D_{(2)}^{X(N)}(\omega)=(g\sqrt{N})^{2}g^{2}\Big[D_{(0)}^{X(N)}(\omega)\Big]^{2}\sum_{m=1}^{M_{g}}\frac{\left\langle \psi_{0}\right|G_{\text{ex}}(\omega)\left|\psi_{m}\right\rangle \left\langle \psi_{m}\right|G_{\text{ex}}(\omega)\left|\psi_{0}\right\rangle }{\omega-((\omega_{\text{ph}}+\beta)+\omega_{g,m})+i\frac{(\kappa + \gamma_\text{vib})}{2}-g^{2}N\frac{|\langle\psi_{0}|\varphi_{0}\rangle|^{2}}{\omega-(\omega_{e}+\omega_{g,m})+i\frac{(\gamma+\gamma_{vib})}{2}}}
\end{align}
\normalsize

\subsection{Stokes cross-section}

\noindent Considering real Franck-Condon factors and $\gamma\to0$, we write $D_{(2)}^{X(N)}(\omega)$
near $\omega=\Omega$, where $\Omega$ is the lowest
peak in the absorption spectra,
\begin{subequations}
\begin{align}
D_{(2)}^{X(N)}(\omega) & \approx(g\sqrt{N})^{2}g^{2}\Big[\frac{1}{\omega-\Omega+i\frac{\Gamma}{2}}\Big]^{2}\sum_{m=1}^{M_{g}}\frac{\Big|\left\langle \psi_{0}\right|G_{\text{ex}}(\omega)\left|\psi_{m}\right\rangle \Big|^{2}}{\omega-(\Omega+\omega_{g,y})+i\frac{\Gamma}{2}}\\
 & \approx(g\sqrt{N})^{2}g^{2}\Big[D_{(0)}^{X(N)}(\omega)\Big]^{2}S_{\text{Raman}}(\omega_{L}=\omega,\omega_{S}=\Omega)
\end{align}
\end{subequations}

\subsection{Conditional anti-Stokes cross-section}

\noindent Considering real Franck-Condon factors and $\gamma\to0$,we write $D_{(2)}^{X(N)}(\omega)$
near $\omega=\Omega$, where $\Omega$ is the lowest
peak in the absorption spectra,

\begin{subequations}
\begin{align}
D_{(2)}^{X(N)}(\omega) & \approx(g\sqrt{N})^{2}g^{2}\Big[\frac{1}{\omega-\Omega+i\frac{\Gamma}{2}}\Big]^{2}\sum_{m=1}^{M_{g}}\frac{\Big|\left\langle \psi_{0}\right|G_{\text{ex}}(\Omega)\left|\psi_{m}\right\rangle \Big|^{2}}{\omega-(\Omega+\omega_{g,y})+i\frac{\Gamma}{2}}\\
 & \approx(g\sqrt{N})^{2}g^{2}\Big[D_{(0)}^{X(N)}(\omega)\Big]^{2}S_{\text{Raman}}^{C}(\omega_{L}=\Omega,\omega_{AS}=\omega)
\end{align}
\end{subequations}

\subsection{Underlying Stokes and anti-Stokes mechanism}

\noindent Thus, $D_{(2)}^{X(N)}(\omega)$ near the lower absorption peak ($\Omega$)
can be expressed in terms of the underlying mechanism involving Stokes
and subsequent anti-Stokes process
\begin{align}
D_{(2)}^{X(N)}(\omega)\approx(g\sqrt{N})^{2}g^{2}\Big[D_{(0)}^{X(N)}(\omega)\Big]^{2}\sqrt{S_{\text{Raman}}(\omega_{L}=\omega,\omega_{S}=\Omega)}\sqrt{S_{\text{Raman}}^{C}(\omega_{L}=\Omega,\omega_{AS}=\omega)}
\end{align}

% \subsection{Absorption via symmetric cavity mode}

% $A^{X}(\omega)$ is the non-trivial part of the absorption owing to the anharmonicities of the molecules. Let $\omega_{\text{ph}}+\beta=\omega_{c}$ and $\kappa_{i}=\frac{\kappa_{ex}}{2}=\frac{\kappa}{4}\implies\kappa_{i}+\frac{\kappa_{ex}}{2}=\frac{\kappa}{2}$. The photon Green's function for arbitary number of molecules coupled to the microtoroidal system that governs this non-trivial part can be computed using continued fraction approach developd by koner et al.

% \begin{align}
%     D_{N}^{XR}(\omega)=\frac{1}{\omega-H_{\text{ph},n}+i\frac{\kappa}{2}-V_{0}\frac{1}{\omega-H_{e,0}+i\frac{\gamma}{2}-v_{0}^{\dagger}\frac{1}{\omega-H_{\text{ph},1}+i\frac{(\kappa+\gamma_{\text{vib}})}{2}-V_{1}^{\dagger}\frac{1}{\frac{\ddots}{\omega-H_{e,N-1}+i\frac{(\gamma+\gamma_{\text{vib}})}{2}-v_{N}^{\dagger}\frac{1}{\omega-H_{\text{ph},N}+i\frac{(\kappa+\gamma_{\text{vib}})}{2}}v_{N}}}V_{1}}}}
% \end{align}

% Here, $H_{\text{ph},n}$ denotes the effective cavity single-photon Hamiltonian for the subspace where n molecules have phonons in the ground state and $H_{e,n}$ corresponds to the photonless subspace with one molecule excited and n other molecules having photons in the ground state. $V_{n}$ is the coupling that drives the collective $\mathcal{O}(\lambda\sqrt{N})$ 'Rayleigh' transitions that conserve the number of molecules with ground state phonons, and $v_{n}$ mediate the single-molecule $\mathcal{O}(\lambda)$ 'Raman' transitions that create/destroy the phonons in one of the molecules.

\begin{table}
	\centering
	% Captions go above tables
	\caption{\textbf{Vibrational mode parameters of isoprene.} 
		This table lists the vibrational mode parameters used to model isoprene, including Franck-Condon factors ($\Delta$) and vibrational frequencies ($\omega$) in electron volts (eV).}
	\label{tab:isoprene_modes} % give each table a logical label name
	
	\begin{tabular}{lcc} % three columns, alignment for each
		\\
		\hline
		Mode & $\Delta$ & $\omega$ (eV) \\
		\hline
		1 & 1.77  & 0.202 \\  
		2 & -0.39 & 0.198 \\  
		3 & 0.36  & 0.181 \\  
		4 & -0.30 & 0.180 \\  
		5 & -0.34 & 0.177 \\  
		6 & -0.35 & 0.172 \\  
		7 & -1.01 & 0.161 \\  
		8 & 0.41  & 0.131 \\  
		9 & 0.37  & 0.123 \\  
		10 & 0.37  & 0.117 \\  
		\hline
	\end{tabular}
\end{table}

% \begin{table}[h]
%     \centering
%     \renewcommand{\arraystretch}{1.5} % Increases vertical spacing
%     \setlength{\tabcolsep}{15pt} % Increases column width
%     \caption{Parameters to model the vibrational modes of isoprene~\cite{Heller_isoprene}}
%     \begin{tabular}{|c|c|}
%         \hline
%         \rule{0pt}{15pt} \textbf{$\Delta$} & \textbf{$\omega$ (eV)} \\ \hline
%         1.77  & 0.202 \\ \hline  
%         -0.39 & 0.198 \\ \hline  
%         0.36  & 0.181 \\ \hline  
%         -0.30 & 0.180 \\ \hline  
%         -0.34 & 0.177 \\ \hline  
%         -0.35 & 0.172 \\ \hline  
%         -1.01 & 0.161 \\ \hline  
%         0.41  & 0.131 \\ \hline  
%         0.37  & 0.123 \\ \hline 
%         0.37  & 0.117 \\ \hline
%     \end{tabular}
%     \label{tab:isoprene_modes}
% \end{table}

\section{Calculation of Light-Matter Coupling Strength ($g$)} \label{sec:g_calculation}

\noindent The single-molecule light-matter coupling strength, $g$, is given by~\cite{heebner2008optical}
\begin{align}
    g = |d_{12}|\sqrt{\frac{(\omega_{\text{ph}}+\beta)}{2\hbar\epsilon_{0}V_{m}}}
\end{align}
\noindent where $|d_{12}|$ is the electronic transition dipole moment of the molecule, $(\omega_{\text{ph}}+\beta)$ is the cavity frequency, $\epsilon_{0}$ is the permittivity of free space, and $V_{m}$ is the cavity mode volume.\\

\noindent For isoprene, the transition dipole moment is calculated using its characteristic transition wavelength, $\lambda_{12} = 205.74$ nm, and oscillator strength, $f_{12} = 0.5523$~\cite{martins2009valence},
\begin{align}
    f_{12} = \frac{2}{3} \frac{m_{e}}{\hbar^{2}} E_{21} |r_{21}|^{2},
\end{align}
\noindent, which gives

\begin{subequations}
\begin{align}
    |d_{12}| &= 1e \times |r_{21}| \\
             &= 1e \times \sqrt{\frac{3\hbar^{2}f_{12}\lambda_{12}}{2m_{e}hc}} \\
             &= 5\text{D}.
\end{align}
\end{subequations}
\noindent The cavity is red-detuned by $0.3$ eV relative to the electronic transition of isoprene, yielding
\begin{align}
    (\omega_{\text{ph}}+\beta) = \omega_{eg} - 0.3\, \text{eV} = 5.726\, \text{eV}.
\end{align}
\noindent The mode volume is taken as six times the diffraction limit for a microtoroidal cavity with refractive index $n = 1.44$:
\begin{align}
    V_{\text{m}} = 6 \times V_{\text{diff}} = 6 \times \left(\frac{\lambda}{n}\right)^{3} = 21.2 \times 10^{-3} \, \mu\text{m}^{3}.
\end{align}
\noindent Substituting these values, the light-matter coupling strength is
\begin{subequations}
\begin{align}
    g &= |d_{12}|\sqrt{\frac{(\omega_{\text{ph}}+\beta)}{2\hbar\epsilon_{0}V_{m}}} \\
      &= 1.6 \times 10^{-4}\, \text{eV}.
\end{align}
\end{subequations}
% \section{Free Spectral Range (FSR)}

% \noindent To avoid interference with neighboring modes, the Free Spectral Range (FSR) must satisfy
% \begin{align}
%     FSR = \omega_{f+1} - \omega_{f} = \frac{c}{nR} = 0.3\, \text{eV}.
% \end{align}

% \noindent This condition is met with a microtoroidal cavity of diameter $D = 1 \, \mu\text{m}$. Thus, the microtoroidal cavity mode that couples with the molecule corresponds to $m = 19$.

\section{Maximum Number of Molecules coupled to a microtoroidal resonator}

\noindent The maximum number of molecules coupled to the resonator is computed using the surface density of the molecules. The surface density of monolayer of receptors is $\rho=5 \times 10^{14}$~\cite{zhuravlev1987concentration}. Thus, the maximum number of molecules coupled to the resonator is

\begin{subequations}
\begin{align}
    N_{\text{max}} &= \rho \times A \\
                   &= \rho \times \pi^{2} \times D_{\lambda_{c}} \times d_{\lambda_{c}} \\
                   &= 5 \times 10^{14} \times 10^{4} \times \pi^{2} \times 10^{-6} \times 0.2 \times 10^{-6} \\
                   &\approx 10^{7} \text{ molecules}
\end{align}   
\end{subequations}
% \textcolor{red}{\section{Application to Terahertz Raman Spectroscopy}
% The IR microtoroidal resonator in Heylman et. al. \cite{randy_nat_photonics} features an FSR of $5$ meV, making it well-suited for detecting low-energy vibrational modes in the terahertz regime. This opens possibilities for enhanced Raman spectroscopy of biologically and chemically relevant molecules, where vibrational features fall within the meV scale.
% Moreover, by leveraging larger microtoroidal resonators, $N_{\text{max}}$ can be significantly increased, further amplifying the Raman signal beyond what is demonstrated in our study. This enhanced interaction strength provides new opportunities for high-sensitivity spectroscopic applications, particularly in label-free molecular detection.
% \textbf{write about Terahertz Raman}}

\section{Appendix: Borel summation}

\noindent Here, we restate the mathematics behind Borel summation for completeness~\cite{hardy2024divergent, borel1899memoire}. Let us consider the following series:
\begin{align}
S=A\sum_{n=0}^{\infty}B^{n}.
\end{align}
We define a slightly different power series in an auxiliary variable
$\lambda$:

\begin{align}
f(\lambda)=A\sum_{n=0}^{\infty}(B\lambda)^{n},
\end{align}
where setting $\lambda=1$ recovers the original sum. The Borel transform
$\hat{f}(t)$ is defined by replacing each coefficient $a_{n}=B^{n}A$
with $\frac{a_{n}}{n!}t^{n}.$ Conversely:

\begin{align}
\hat{f}(t)=\sum_{n=0}^{\infty}\frac{a_{n}}{n!}t^{n}=\sum_{n=0}^{\infty}\frac{B^{n}A}{n!}t^{n}=A\sum_{n=0}^{\infty}\frac{B^{n}}{n!}t^{n}=Ae^{Bt}.
\end{align}
The Borel sum $\mathcal{B}[f(\lambda)]$ of $f(\lambda)$ is then
the Laplace-type integral

\begin{align}
\mathcal{B}[f(\lambda)]=\int_{0}^{\infty}e^{-u}\hat{f}(\lambda u)du.
\end{align}
In our case, 

\begin{align}
\hat{f}(\lambda u)=Ae^{B(\lambda u)},
\end{align}
so

\begin{align*}
\mathcal{B}[f(\lambda)] & =\int_{0}^{\infty}e^{-u}Ae^{B(\lambda u)}du,\\
 & =A\int_{0}^{\infty}e^{u(B\lambda-1)}du.
\end{align*}
The original sum $S$ corresponds to $\lambda=1$, so we set $\lambda=1$:

\begin{align}
S_{\text{Borel }}=A\int_{0}^{\infty}e^{u(B-1)}du.
\end{align}
Now if $\Re(B)<1$ the exponential $e^{u(B-1)}$ decays for $u>0$,
so the integral converges and equals $\frac{A}{1-B}$. If $\Re(B)\ge1$:
then $\Re(B-1)\ge0$ and the real-axis integral $\int_{0}^{\infty}e^{u(B-1)}du$
diverges. However, there are complex analysis tricks to analytically
continue to the space of $\Re(B)<1$. Rather than integrating along
the positive real axis $t\in[0,\infty)$, one rotates the integration
contour into the complex plane in such a way that the real part of
the exponent becomes negative, ensuring convergence. 

\medskip{}
\noindent Let $M=B-1.$ If $\Re(M)>0$, the real-axis integral ${\it \int_{0}^{\infty}e^{Mt}dt}$
diverges. However, consider integrating along the line $t=re^{i\theta}$
for $r$ from $0$ to $\infty$, where $\theta$ is chosen so that, 

\begin{align}
\Re[M(re^{i\theta})]=r\Re[Me^{i\theta}]\quad<0,
\end{align}
ensuring the exponential decays at large $r$. Concretely, if $M$
is positive real number ($i.e.$ $B>1$ real), you can take $\theta=\pi$
(which amounts to integration along the negative real axis). Then

\begin{align}
t=re^{i\pi}=-r,\quad r\in[0,\infty),
\end{align}
so $\Re(M\cdot(-r))=-r\Re(M)<0$. That yields convergence. Along the
line $t=re^{i\theta},$ we have $dt=e^{i\theta}dr$. So formally:

\begin{align}
\int_{0}^{\infty}e^{(B-1)t}dt\rightarrow\int_{0}^{\infty}e^{(B-1)re^{i\theta}}e^{i\theta}dr.
\end{align}
If we choose $\theta$ so that $\Re((B-1)e^{i\theta})<0$ (e.g. $\theta=\pi$
when $B>1$ and is purely real), the integral converges. We then have, 

\begin{align}
\int_{0}^{\infty}e^{(B-1)t}dt=\int_{C}e^{(B-1)t}dt,
\end{align}
provided there are no singularities or branch cuts crossing the path.
In fact, it turns out that the integral value is the same as if we had
$\Re(B)<1$. One recovers, 

\begin{align}
S_{\text{Borel }}=A\int_{\mathbb{\mathbb{C}}}e^{(B-1)t}dt=\frac{A}{1-B},
\end{align}
which is the same answer as one would get without the analytic continuation.

\bibliography{references}% Produces the bibliography via BibTeX.
\bibliographystyle{sciencemag.bst}

%---------------------X------------------------X------------------------
% SUPPLEMENTARY MATERIALS